\documentclass[twocolumn,trackchanges]{aastex631}

\usepackage{gensymb}

\begin{document}


\title{Characterizing the Broadband Reflection Spectrum of MAXI~J1803-298 During its 2021 Outburst with \textit{NuSTAR} and \textit{NICER}}

\author[0000-0002-5966-4210]{Oluwashina K. Adegoke}
\affiliation{Cahill Center for Astronomy \& Astrophysics, California Institute of Technology,
Pasadena, CA 91125, USA}

\author[0000-0003-3828-2448]{Javier A. Garc\'ia}
\affiliation{X-ray Astrophysics Laboratory, NASA Goddard Space Flight Center, Greenbelt, MD 20771, USA}
\affiliation{Cahill Center for Astronomy \& Astrophysics, California Institute of Technology,
Pasadena, CA 91125, USA}

\author[0000-0002-8908-759X]{Riley M. T. Connors}
\affiliation{Department of Physics, Villanova University, 800 E. Lancaster Avenue, Villanova, PA 19085, USA}

\author[0000-0002-5770-2666]{Yuanze Ding}
\affiliation{Cahill Center for Astronomy \& Astrophysics, California Institute of Technology,
Pasadena, CA 91125, USA}

\author[0000-0003-4216-7936]{Guglielmo Mastroserio}
\affiliation{Dipartimento di Fisica, Universit\`a Degli Studi di Milano, Via Celoria, 16, Milano, 20133, Italy}

\author[0000-0002-5872-6061]{James F. Steiner}
\affiliation{Harvard-Smithsonian Center for Astrophysics, 60 Garden Street, Cambridge, MA 02138, USA}

\author[0000-0002-5311-9078]{Adam Ingram}
\affiliation{School of Mathematics, Statistics and Physics, Newcastle University, Newcastle upon Tyne NEI 7RU, UK}

\author[0000-0003-2992-8024]{Fiona A. Harrison}
\affiliation{Cahill Center for Astronomy \& Astrophysics, California Institute of Technology,
Pasadena, CA 91125, USA}

\author[0000-0001-5506-9855]{John A. Tomsick}
\affiliation{Space Sciences Laboratory, 7 Gauss Way, University of California, Berkeley, CA 94720-7450, USA}

\author[0000-0003-0172-0854]{Erin Kara}
\affiliation{MIT Kavli Institute for Astrophysics and Space Research, MIT, 70 Vassar Street, Cambridge, MA 02139, USA}

\author[0000-0002-4992-4664]{Missagh Mehdipour}
\affiliation{Space Telescope Science Institute, 3700 San Martin Drive, Baltimore, MD 21218, USA}


\author[0000-0001-5709-7606]{Keigo Fukumura}
\affiliation{Department of Physics and Astronomy, James Madison University, 800 South Main Street, Harrisonburg, VA 22807, USA}

\author[0000-0003-2686-9241]{Daniel Stern}
\affiliation{Jet Propulsion Laboratory, California Institute of Technology, Pasadena, CA 91109, USA}

\author[0000-0002-6159-5883]{Santiago Ubach}
\affiliation{Harvard-Smithsonian Center for Astrophysics, 60 Garden Street, Cambridge, MA 02138, USA}

\author[0000-0002-2235-3347]{Matteo Lucchini}
\affiliation{Anton Pannekoek Institute for Astronomy, University of Amsterdam, Science Park 904, 1098 XH Amsterdam, The Netherlands}






\begin{abstract}
\rm{MAXI~J1803-298} is a transient black hole candidate discovered in May of 2021 during an outburst that lasted several months. Multiple X-ray observations reveal recurring ``dipping'' intervals in several of its light curves, particularly during the hard/intermediate states, with a typical recurrence period of $\sim7\,\mathrm{hours}$. We report analysis of four \textit{NuSTAR} observations of the source, supplemented with \textit{NICER} data where available, over the duration of the outburst evolution covering the hard, intermediate and the soft states. Reflection spectroscopy reveals the black hole to be rapidly spinning
($a_*=0.990\pm{0.001}$) with a near edge-on viewing angle 
($i=70\pm{1}{\degree}$).
Additionally, we show that the light-curve dips are caused by photo-electric absorption from a moderately ionized absorber whose origin is not fully understood, although it is likely linked to material from the companion star impacting the outer edges of the accretion disk. 
We further detect absorption lines in some of the spectra, potentially associated with {Fe~\sc{xxv}} and {Fe~\sc{xxvi}}, indicative of disk winds with moderate to extreme velocities. 
During the intermediate state and just before transitioning into the soft state, the source showed a sudden flux increase which we found to be dominated by soft disk photons and consistent with the filling of the inner accretion disk, at the onset of state transition. In the soft state, we show that models of disk self-irradiation provide a better fit and a preferred explanation to the broadband reflection spectrum, consistent with previous studies of other accreting sources. 

\end{abstract}



\section{Introduction} \label{sec:one}
While it is generally believed that $10^{7}-10^{9}$ stellar-mass black holes lurk within the Milky Way \citep[e.g.,][]{2022ApJ...933L..23L}, only a few dozens have been detected, many of those during an outburst when the source suddenly becomes very bright in X-rays \citep{2016ApJS..222...15T}. This is usually caused by an abrupt increase in accretion rate resulting in an increase in the transient black hole's luminosity by several orders of magnitude, typically in the range $\sim10^{34}-10^{38}\,\mathrm{erg\,s^{-1}}$. At the onset of an outburst, a black hole X-ray binary (BHXB) rises in the so-called hard state with the X-ray spectrum dominated by a power law believed to be produced from the inverse Compton scattering of seed disk photons in a hot ($T\sim10^{8}-10^{9}\,\mathrm{K}$), moderately optically thick ($\tau\sim1-2$) ``corona'' \citep{1993ApJ...413..680H, 1993ApJ...413..507H, 1997ApJ...487..759D, 2003MNRAS.342..355Z}. The source usually transitions into the intermediate state near the peak of the outburst. In this state, alongside the non-thermal power-law continuum, a thermal disk component also becomes apparent. As the source progresses into the soft state, the thermal disk component becomes dominant \citep[for a review, see e.g.,][]{2006ARA&A..44...49R, 2016ASSL..440...61B, 2022hxga.book....9K}. These spectral state changes are generally associated with evolution of the accretion geometry. 

In the soft state, it is generally accepted that the accretion disk extends all the way to the innermost stable circular orbit (ISCO) of the black hole \citep{1997ApJ...489..865E, 2005ApJ...624..295H, 2005A&A...440..207B, 2007A&ARv..15....1D, 2010ApJ...718L.117S}.
The extent of the disk in the hard state is however still being debated, especially for luminosities in the moderate range of $0.1-10\%$ of the Eddington limit. While the disk appears to be truncated for some systems in this state \citep[e.g.,][]{2015ApJ...813...84G, 2016MNRAS.458.2199B, 2021ApJ...906...69Z}, other sources suggest an untruncated disk extending down to the ISCO \citep[e.g.,][]{2006ApJ...653..525M, 2008MNRAS.387.1489R, 2022ApJ...935..118C}. Black holes have also been known to show persistent and steady jets, parsec-scale ballistic jets and quasi-periodic oscillations (QPOs) during a single outburst cycle \citep[e.g.,][]{2004MNRAS.355.1105F, 2022hxga.book....9K}.

A prominent feature that is usually imprinted on the broadband X-ray spectrum of BHXBs is the reflection component. This is caused by the interaction of hard X-ray photons with the disk, producing the broad $\rm{Fe\,K{\alpha}}$ line between $6-7\,\mathrm{keV}$ \citep{1989MNRAS.238..729F, 1991ApJ...376...90L, 1988ApJ...335...57L} and the Compton reflection hump which peaks around $20\,\mathrm{keV}$. Because of the strong general relativistic (GR) effects experienced by matter in the immediate vicinity of a black hole, the profile of the reflection spectrum, most importantly the $\rm{Fe\,K{\alpha}}$ line, is significantly distorted. The degree of this distortion provides a useful way to quantify the properties of the accretion environment around the black hole, including the geometry, composition, inclination, location of the inner radius and, crucially, the black hole spin \citep{2010MNRAS.409.1534D, 2014SSRv..183..277R}.

Mass outflows in the form of winds or jets play an important role in the study of accretion processes around black holes, although a detailed understanding of how black hole accretion disks drive winds and jets still remains elusive. For sources observed close to the disk plane, the presence of highly ionized narrow absorption lines from iron 
are known to be signatures of powerful outflowing disk winds \citep[e.g.,][]{2006ApJ...646..394M, 2008ApJ...680.1359M, 2009Natur.458..481N, 2012MNRAS.422L..11P, 2012ApJ...746L..20K, 2014ApJ...784L...2K, 2015ApJ...799L...6M, Neilsen2023}. Around BHXBs, the most important of these lines are He-like {Fe~\sc{xxv}} and H-like {Fe~\sc{xxvi}}, because they can be prominent in very hot and ionized environments \citep[see e.g.,][]{2001ApJS..134..139B}, making them good tracers of winds in regions closest to the black hole where they are launched. These disk winds are observed to be dense and to have typical velocities of $\sim1000\,\mathrm{km\,s^{-1}}$ or less projected along our line of sight. For a number of sources, the presence of disk winds have been shown to be anti-correlated with the onset of relativistic jets \citep[e.g.,][]{2006ApJ...646..394M, 2009Natur.458..481N, 2012ApJ...759L...6M}. In these systems, it is believed that most of the momentum in the jet, typically seen in the low/hard state, is supplanted by wind outflow in the high/soft state.

Flux variation on a wide range of time scales is ubiquitous in both stellar and supermassive black holes and is believed to be connected to the accretion flow process in the regions around the central source \citep[e.g.,][]{1997ARA&A..35..445U, 2006csxs.book...39V, 2010LNP...794..203M, 2019ApJ...870L..13A}. 

\rm{MAXI~J1803-298} 
was detected by \textit{MAXI} \citep{2009PASJ...61..999M} on May 1, 2021 during the onset of its only known outburst \citep{2021ATel14587....1S, 2021ATel14627....1S}. Through follow-up observations, 
the source was observed to show signatures of an accreting black hole while rising in the hard state \citep{2021ATel14602....1B, 2021ATel14597....1B, 2021ATel14606....1H, 2021ATel14609....1X}. The source showed dips in several of its X-ray light curves with a recurrence period of $\sim7\,\mathrm{hours}$ while mostly in the hard/intermediate states \citep{2021ATel14606....1H, 2021ATel14609....1X, 2022MNRAS.511.3922J}. Also while in the hard/intermediate states, QPOs were detected on short time-scales ranging from $0.13\,\mathrm{Hz}$ to $7.61\,\mathrm{Hz}$ \citep{2021ATel14602....1B, 2021ATel14609....1X, 2021ATel14660....1U, 2021ATel14613....1W, 2021ATel14630....1C, 2021ATel14629....1J, 2022MNRAS.511.3922J, 2022ApJ...933...69C, 2023ApJ...949...70C}. The source additionally showed spectral absorption lines, both in the X-ray and optical, consistent with the presence of disk winds \citep{2021ATel14650....1M, 2022ApJ...926L..10M, 2024arXiv240210315Z} and, it has been argued to be rapidly spinning with an inclination close to edge-on \citep{2022ApJ...933...69C, 2022MNRAS.516.2074F, 2023ApJ...949...70C}. \citet{2022ApJ...926L..10M} estimated a conservative mass for the black hole in the system to be in the range $\sim3-10\,\mathrm{M_{\odot}}$ based on optical spectroscopy.

In this paper, we follow the broadband spectral evolution of \rm{MAXI~J1803-298} at different epochs throughout its 2021 outburst, using data from both \textit{NuSTAR} \citep{2013ApJ...770..103H} and \textit{NICER} \citep{2016SPIE.9905E..1HG}. The aim is to characterize and constrain the accretion flow properties as the source goes through different states over the course of the entire outburst. 

The paper is structured as follows. In Section \ref{sec:two}, we present the observations and the data reduction procedure. In Section \ref{sec:three}, we describe the data analysis and the results. In Section \ref{sec:four}, we discuss the implications of the results obtained for all four epochs and we summarize our main conclusions in Section \ref{sec:five}. 
\begin{table*}
\caption{Log of \textit{NuSTAR} and \textit{NICER} observations of \rm{MAXI~J1803-298} used.} 
\centering 
\begin{tabular}{l l l l l} 
\hline\hline 
\textbf{Epoch} & \textbf{Mission} & \textbf{ObsID} & \textbf{Obs. start (UTC)} & \textbf{Exp. (ks)} \\ [0.5ex] 
\hline\hline 
1 & \textit{NuSTAR}& 90702316002 & 2021-05-05 16:46:09 & 27\\
& \textit{NICER}& 4202130104 & 2021-05-05 01:20:30 & 4\\
2 & \textit{NuSTAR}& 80701332002 & 2021-05-14 23:01:09 & 32\\
3 & \textit{NuSTAR}& 90702318002 & 2021-05-23 16:11:09 & 13\\
& \textit{NICER}& 4202130110 & 2021-05-23 09:36:56 & 8\\
& \textit{NICER}& 4202130111 & 2021-05-23 23:33:54 & 3\\
4 & \textit{NuSTAR}& 90702318003 & 2021-06-17 19:46:09 & 16\\
& \textit{NICER}& 4675020124 & 2021-06-17 09:08:09 & 3\\
\hline 
\hline 

\end{tabular}
\label{tab:log} 
\end{table*}
\section{Observations and Data Reduction} \label{sec:two}
\subsection{NuSTAR} \label{subsec:two-one}
\rm{MAXI~J1803-298} was observed by \textit{NuSTAR} on five occasions during its 2021 outburst. The first (Epoch 1) and the fourth (Epoch 4) observations caught the source in the rising hard and the declining soft states, respectively, while the second and the third observations (Epochs 2 \& 3) caught the source in the soft intermediate state (see Table \ref{tab:log}). The source was hardly detected during the fifth observation and is therefore not reported here.

The data were reduced using the standard pipeline Data Analysis Software (\textsc{nustardas}, v.2.1.2) and \textsc{caldb} v20220118. Event files and images were generated with the \texttt{nupipeline} command. In all cases, the source was extracted from a circular region of radius 150'' while the background was extracted from a source-free region of the same radius --- extending to adjacent detectors in some cases. Source and background spectra and light curves, with instrumental responses, were generated using the \texttt{nuproducts} task. For Epochs 1 and 3, spectra were analyzed in the complete $3-79\,\mathrm{keV}$ band while for Epoch 4, spectra were analyzed only in the energy range $3-25\,\mathrm{keV}$ as background tended to dominate above $\sim25\,\mathrm{keV}$ in that observation. For Epoch 2, there is a pronounced discrepancy in the spectra between FPMA and FPMB below $4\,\mathrm{keV}$ (see Fig. \ref{fig:mo_refl}), also reported by \citet{2023ApJ...949...70C}. It is believed to be caused by a rip in the multi layer insulation (MLI) associated with FPMA \citep[see e.g.,][]{2020arXiv200500569M}. 
Because of this, as well as significant background contribution above $70\,\mathrm{keV}$, data below $4\,\mathrm{keV}$ and above $70\,\mathrm{keV}$ were excluded in the spectral analysis of this observation reported in Table \ref{tab:mo_table}.


\subsection{NICER} \label{subsec:two-two}
\rm{MAXI~J1803-298} was observed several times by \textit{NICER} during its 2021 outburst of which a few were simultaneous or quasi-simultaneous with \textit{NuSTAR} and are therefore used for broadband spectral analysis (see Table \ref{tab:log}). 

The data reduction followed standard procedure as outlined in the \textit{NICER} Data Analysis Thread\footnote{\url{https://heasarc.gsfc.nasa.gov/docs/nicer/analysis_threads/}} using \textsc{nicerdas} version 10a. We generated cleaned event files using the {\tt nicerl2} task. {\tt nicerl3-lc} and {\tt nicerl3-spec} were employed to generate light curves and spectra, respectively, with backgrounds --- as well as their associated response files. For spectral background, we chose the \textsc{scorpeon} model with the file output format. The \textit{NICER} data were considered in the energy range $0.3-10\,\mathrm{keV}$ for spectral analysis.


\section{Data Analysis and Results} \label{sec:three}
\begin{figure}
\includegraphics[width=.50\textwidth, angle=0]{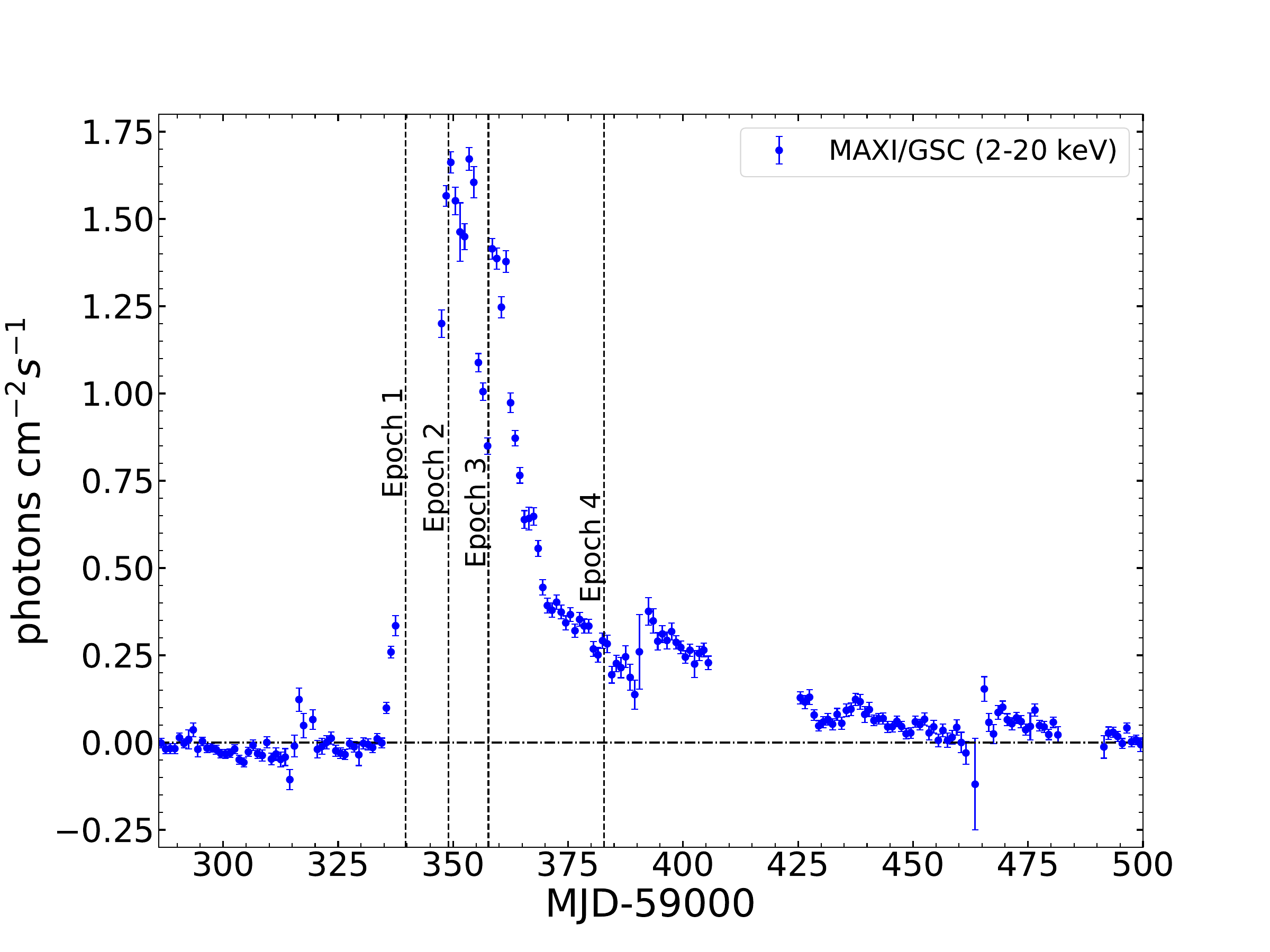}
\caption{The \textit{MAXI} per-day light curve of \rm{MAXI~J1803-298} showing the period corresponding to the entire outburst with the dashed vertical lines being the times of the four \textit{NuSTAR} observations reported.} 
\label{fig:lc_nustar-maxi}
\end{figure}

\begin{figure*}
\includegraphics[width=.5\textwidth, angle=0, trim={1.0cm 0cm 1.5cm 0.5cm}]{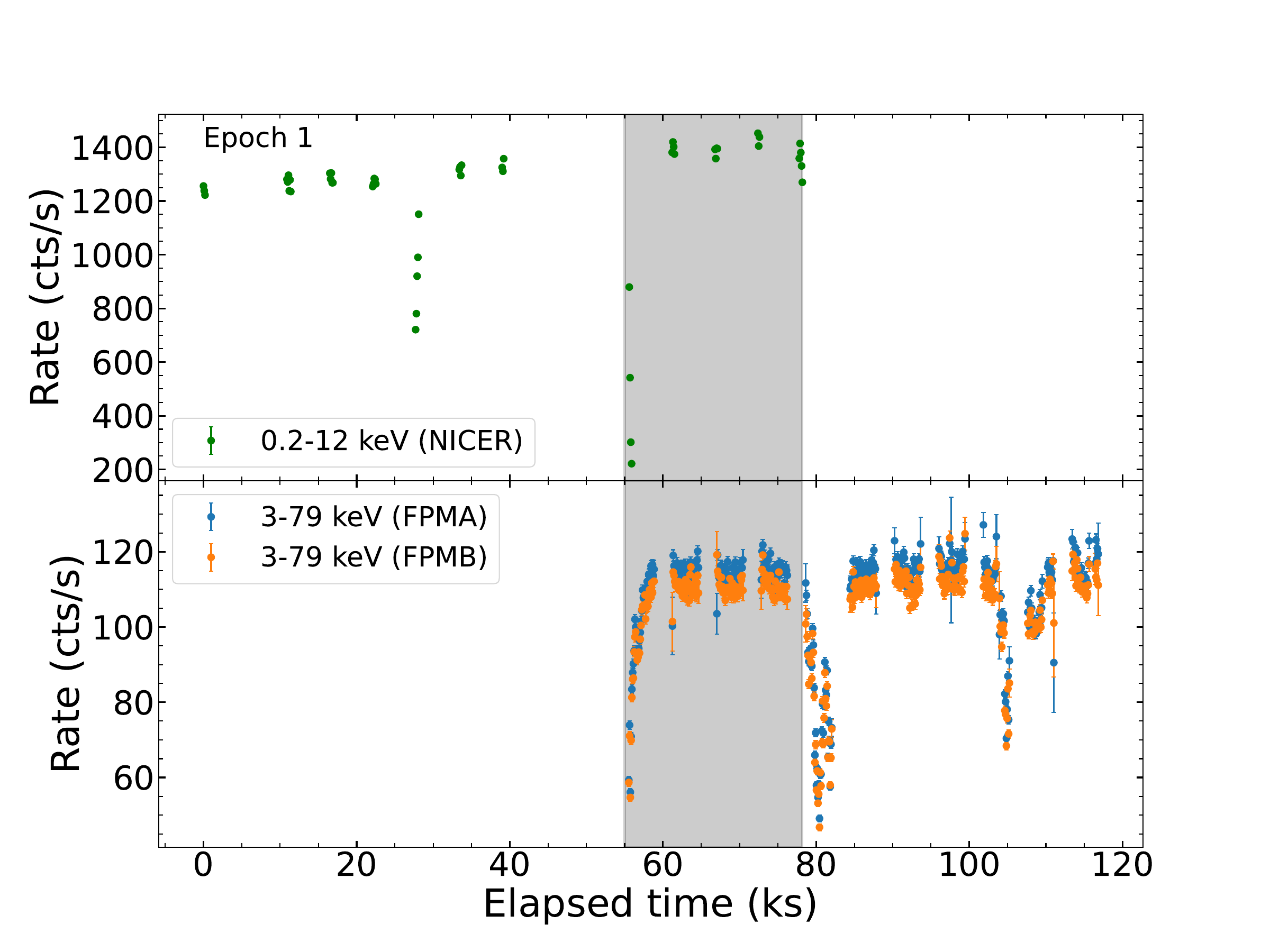}
\includegraphics[width=.5\textwidth, angle=0, trim={1.0cm 0cm 1.5cm 0.5cm}]{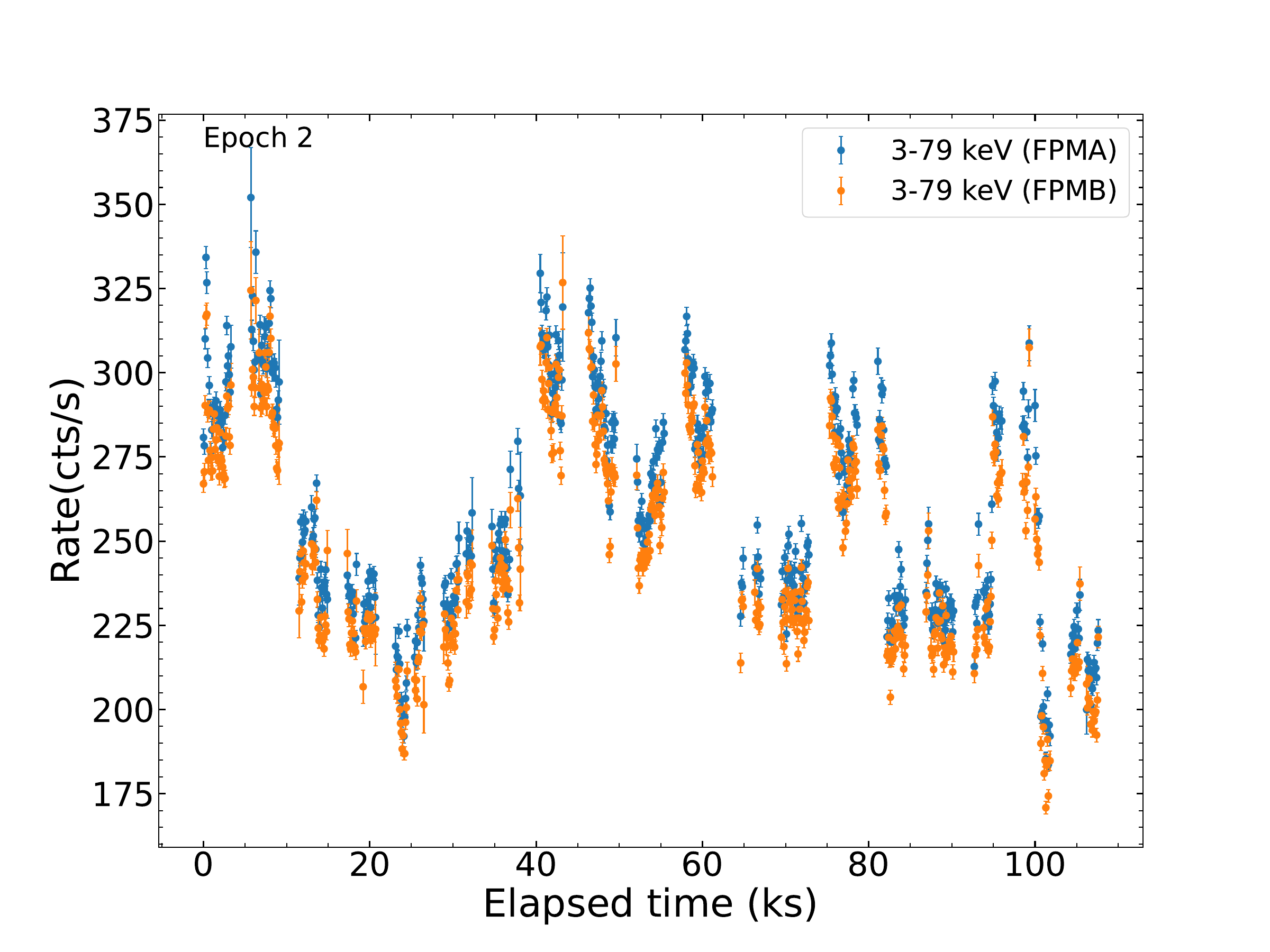}\\
\includegraphics[width=.5\textwidth, angle=0, trim={1.0cm 0cm 1.5cm 0.5cm}]{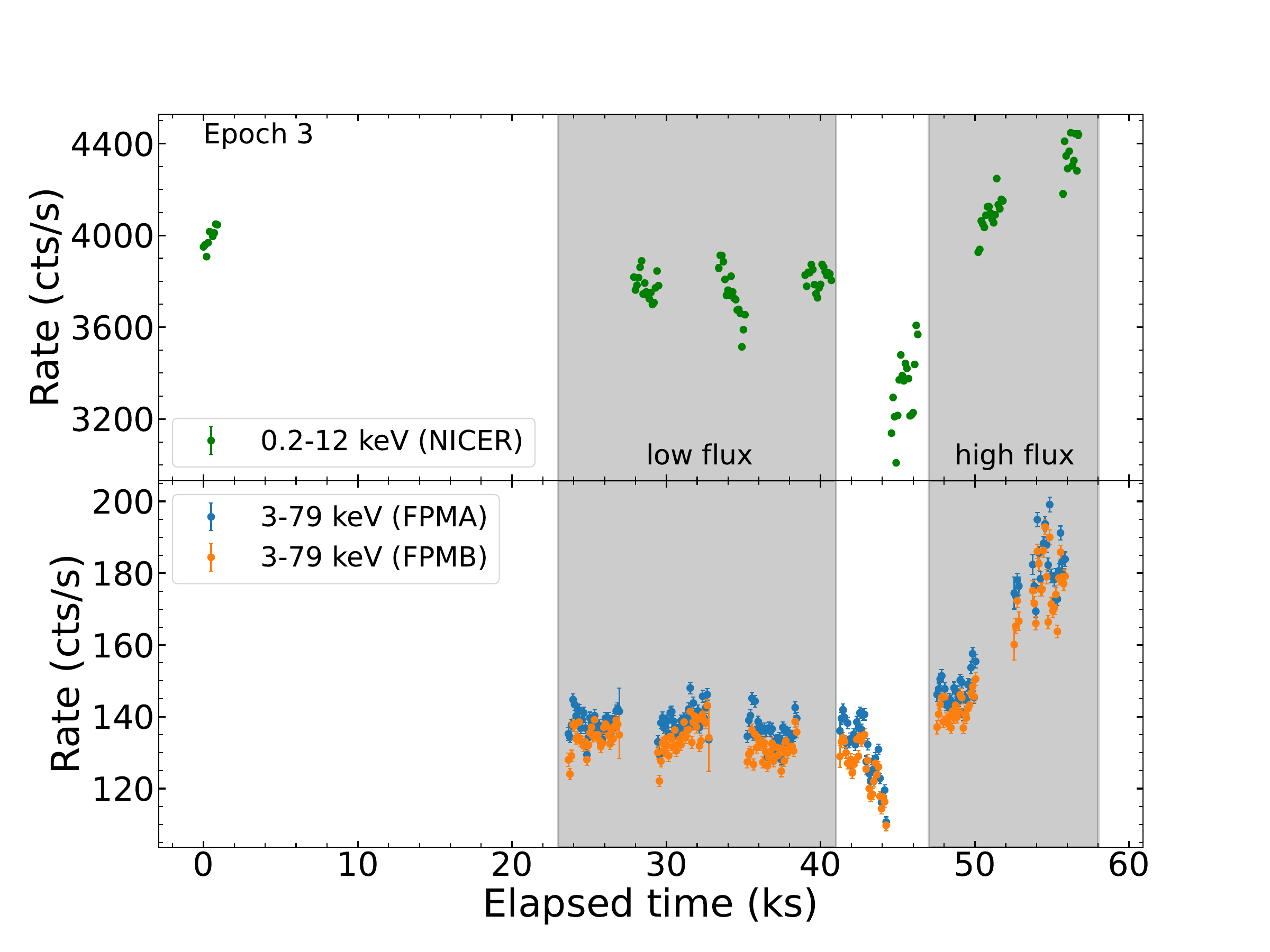}
\includegraphics[width=.5\textwidth, angle=0, trim={1.0cm 0cm 1.5cm 0.5cm}]{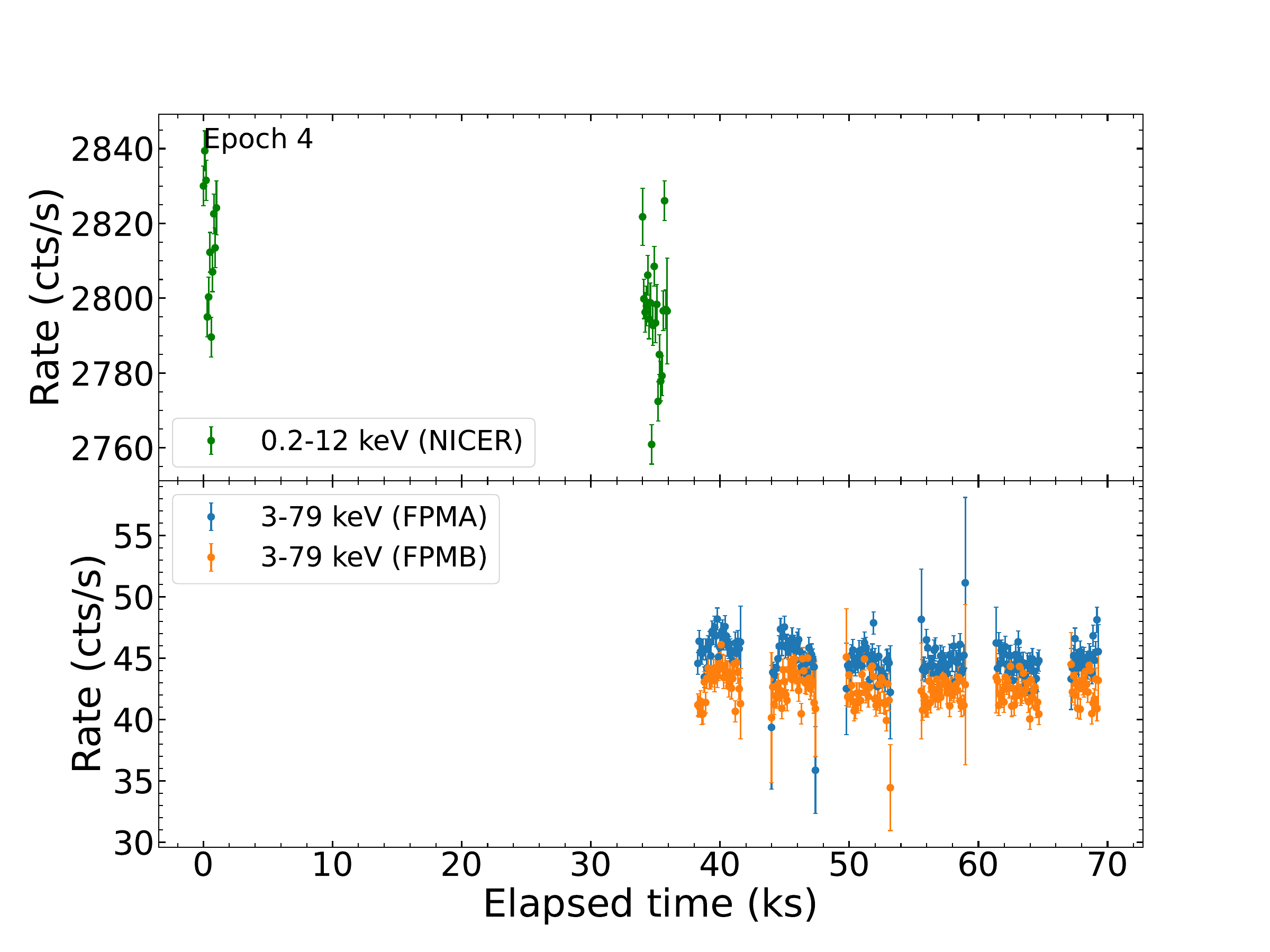}
\caption{\textit{NICER} and \textit{NuSTAR} light curves for Epochs~$1-4$ (note that Epoch~2 does not have concurrent \textit{NICER} observations). All the light curves have time bins of $100\,\mathrm{s}$. The shaded parts of Epoch 1 and Epoch 3 light curves indicate the regions used for the joint spectral fitting described in Section \ref{subsubsec:three-one-two} and Section \ref{subsec:three-three}.}
\label{fig:lc_nustar-nicer}
\end{figure*}

\begin{figure}
\includegraphics[width=.50\textwidth, angle=0, trim={2cm 1cm 2cm 1cm}]{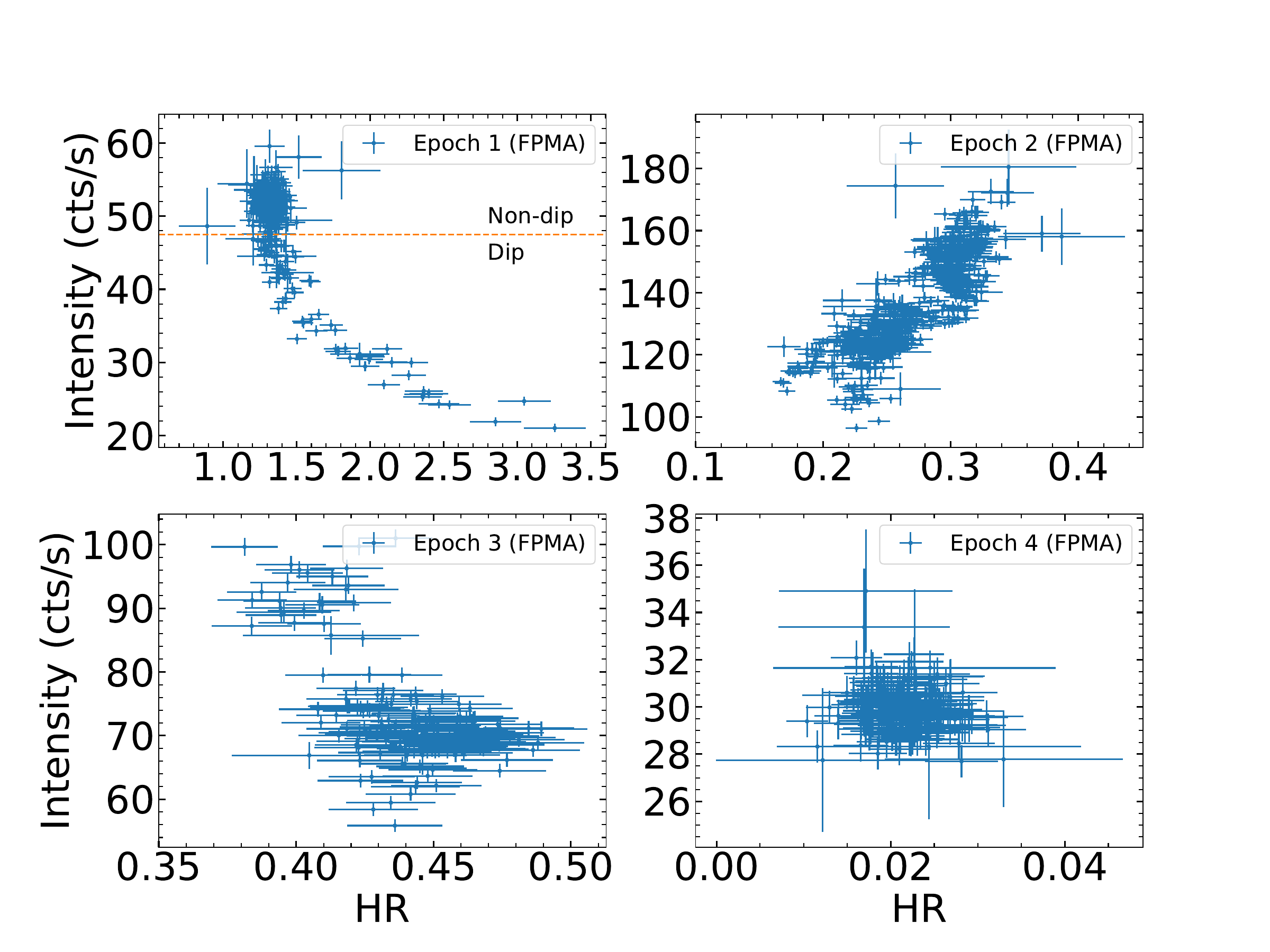}
\caption{HID from FPMA for all four \textit{NuSTAR} epochs. While HR is the ratio of count rates in the hard band ($8.5-15\,\mathrm{keV}$) to those in the soft band ($3-4.5\,\mathrm{keV}$), intensity is the sum of count rates in both bands.
For Epoch 1, the dashed (orange) horizontal line separates data from the non-dip vs. the dip intervals.} 
\label{fig:hid_nu}
\end{figure}

Figure \ref{fig:lc_nustar-maxi} shows the \textit{MAXI}/GSC light curve of \rm{MAXI~J1803-298} covering the periods before, during and after its 2021 outburst. The dashed vertical lines represent the times when the four \textit{NuSTAR} observations were carried out. The individual \textit{NuSTAR} and \textit{NICER} light curves for all epochs (except for Epoch 2 which does not have \textit{NICER} observations) are shown in Fig. \ref{fig:lc_nustar-nicer}. The hardness-intensity diagram (HID) from \textit{NuSTAR} corresponding to each of the epochs is shown in Fig. \ref{fig:hid_nu}. Here, the hardness ratio (HR) is defined as the ratio of count rates in the $8.5-15\,\mathrm{keV}$ range to that in the $3-4.5\,\mathrm{keV}$ range, while the intensity is the sum of count rates in both energy bands. This choice of energy range is made to exclude regions where the reflected spectrum contributes significantly. 

For spectral analysis, the \textit{NuSTAR} data are grouped with a minimum of 40 counts per spectral bin using the ``optmin'' flag in \texttt{ftgrouppha} \citep{2016A&A...587A.151K} to ensure sufficient counts in each spectral bin for the reliable use of $\chi^{2}$ statistics. The same optimal binning scheme is applied to the \textit{NICER} data (implemented by default in the \texttt{nicerl3-spec} task), but without the requirement of a minimum number of counts per bin, given the much high number of \textit{NICER} counts.
All the fits and statistical analyses are performed in \texttt{XSPEC} v12.13.0c \citep{1996ASPC..101...17A}. We model photo-electric absorption in the interstellar medium along the line of sight to the source with \texttt{tbabs} or \texttt{tbfeo}, using the cosmic abundances of \citet{2000ApJ...542..914W} and the cross-sections of \citet{1996ApJ...465..487V}. 
We fix the hydrogen column density to $N_{\rm{H}}=3.2\times10^{21}\,\mathrm{cm^{-2}}$ based on earlier reported best-fit values \citep[e.g.,][]{2021ATel14602....1B, 2021ATel14606....1H}. We also included cross-normalization constants to account for differences in absolute flux calibration between instruments/telescopes. 
In all cases, errors are computed at the 90\% confidence interval for one interesting parameter. For the values in Tables \ref{tab:mo_table} and \ref{tab:relxillns}, this is implemented in \texttt{XSPEC} with Monte Carlo Markov Chain (MCMC) using the Goodman-Weare algorithm. We use 50 walkers and a total chain length of $2\times10^{6}$ for each run, with the first $10^{6}$ steps discarded. These values were chosen after a number of trials and using the plot of the MCMC statistic against chain step to ensure convergence of the chains for all epochs. 

\begin{figure}
\includegraphics[width=.50\textwidth, angle=0, trim={3cm 1cm 1cm 1cm}]{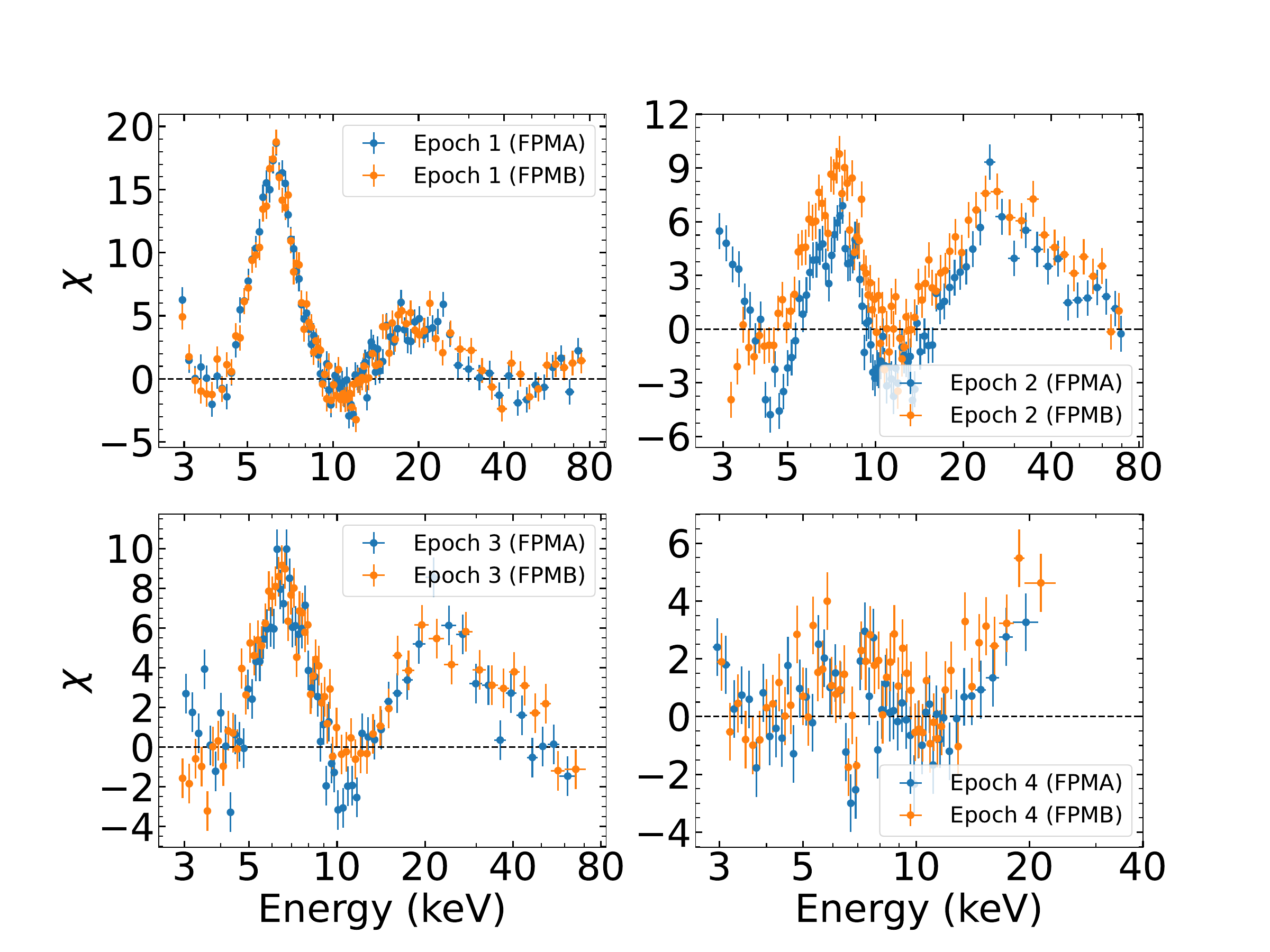}
\caption{Residuals for all four epochs based on an absorbed \texttt{diskbb+cutoffpl} fit to the \textit{NuSTAR} data in the energy range $3-4.5\,\mathrm{keV}$, $8.5-15\,\mathrm{keV}$ and $40-79\,\mathrm{keV}$ ($40-70\,\mathrm{keV}$ for Epoch 2), extrapolated over the complete energy band (the complete energy band for Epoch 4 is $3-25\,\mathrm{keV}$). The spectra have been rebinned for plotting purposes.} 
\label{fig:mo_refl}
\end{figure}

We started by fitting phenomenological models to the \textit{NuSTAR} spectra including a \texttt{diskbb} and a \texttt{cutoffpl} for all four epochs in the energy range $3-4.5\,\mathrm{keV}$,
$8.5-15\,\mathrm{keV}$ and $40-79\,\mathrm{keV}$ (except for Epochs 2 and 4 which are considered only up to $70\,\mathrm{keV}$ and $25\,\mathrm{keV}$ respectively). 
This is done to exclude the energy range corresponding to the prominent reflection features, i.e., the broad Fe K$\alpha$ line and the Compton hump. 
 When extrapolated to include the complete energy band, strong, broad reflection features are evident in all epochs. These include the Fe K${\alpha}$ complex at $\sim6.4\,\mathrm{keV}$ and the Compton hump peaking around $20\,\mathrm{keV}$ --- shown in Fig. \ref{fig:mo_refl}. To characterize the reflection features exhibited by the source and to probe the evolution of the accretion flow over the course of the outburst, we employed the \texttt{relxillCp} flavor of the state-of-the-art reflection model \texttt{relxill} \citep{2014MNRAS.444L.100D, 2014ApJ...782...76G} --- analyzing each epoch separately first and then jointly, subsequently. In all cases, we set the inner radius $R_{in}$ at the ISCO and fit for the spin since these parameters are degenerate.

For cases where Fe K absorption features are detected in the spectra, we carried out a rigorous Monte Carlo test to determine the significance of these lines. The approach is described in Section \ref{subsec:three-six}.

For the analysis reported in Tables \ref{tab:mo_table} and \ref{tab:relxillns}, we replaced \texttt{tbabs} with \texttt{tbfeo} to compensate for the known low-energy residuals in \textit{NICER}, allowing both iron and oxygen abundances to be free in all cases when \textit{NICER} data is included. 

\subsection{Epoch 1: Persistent vs. Dip Spectra} \label{subsec:three-one}
\subsubsection{NuSTAR} \label{subsubsec:three-one-one}
This observation, from May 5, was carried out during the rising phase of the outburst while the source was still in the hard state (see Fig. \ref{fig:lc_nustar-maxi}). FPMA and FPMB observations were for durations of $26.5\,\mathrm{ks}$ and $26.8\,\mathrm{ks}$, respectively. The source shows recurrent light-curve dips during this observation that might be indicative of some form of obscuration --- intermittently seen in high inclination systems. Spectral analysis based on \textit{AstroSat} data showing similar flux dips from a later observation carried out between May 11 and 12 are reported in \citet{2022MNRAS.511.3922J}. \textit{AstroSat} \citep{2006AdSpR..38.2989A, 2014SPIE.9144E..1SS} observed the source when it was undergoing a transition from the hard-intermediate state to the
soft-intermediate state.

To probe the nature and cause of the dips, we generated spectra for the dip intervals (hereafter ``dip'' spectra) and for intervals without the dips (hereafter ``persistent'' spectra). The individual good time intervals (GTIs) were generated in {\tt xselect} using the cleaned event files from {\tt nupipeline}. The new GTIs were then employed to create spectra for the persistent and dip intervals. The accumulated durations for the persistent spectra are $19.5\,\mathrm{ks}$ for FPMA and $19.4\,\mathrm{ks}$ for FPMB. To obtain a higher signal-to-noise ratio, we combine spectra from all three \textit{NuSTAR} dips shown in the top left plot of Fig. \ref{fig:lc_nustar-nicer}, which amount to $6.7\,\mathrm{ks}$ for FPMA and $7.0\,\mathrm{ks}$ for FPMB.

\begin{figure*}
\includegraphics[width=.50\textwidth, angle=0, trim={3cm 1cm 2cm 1cm}]{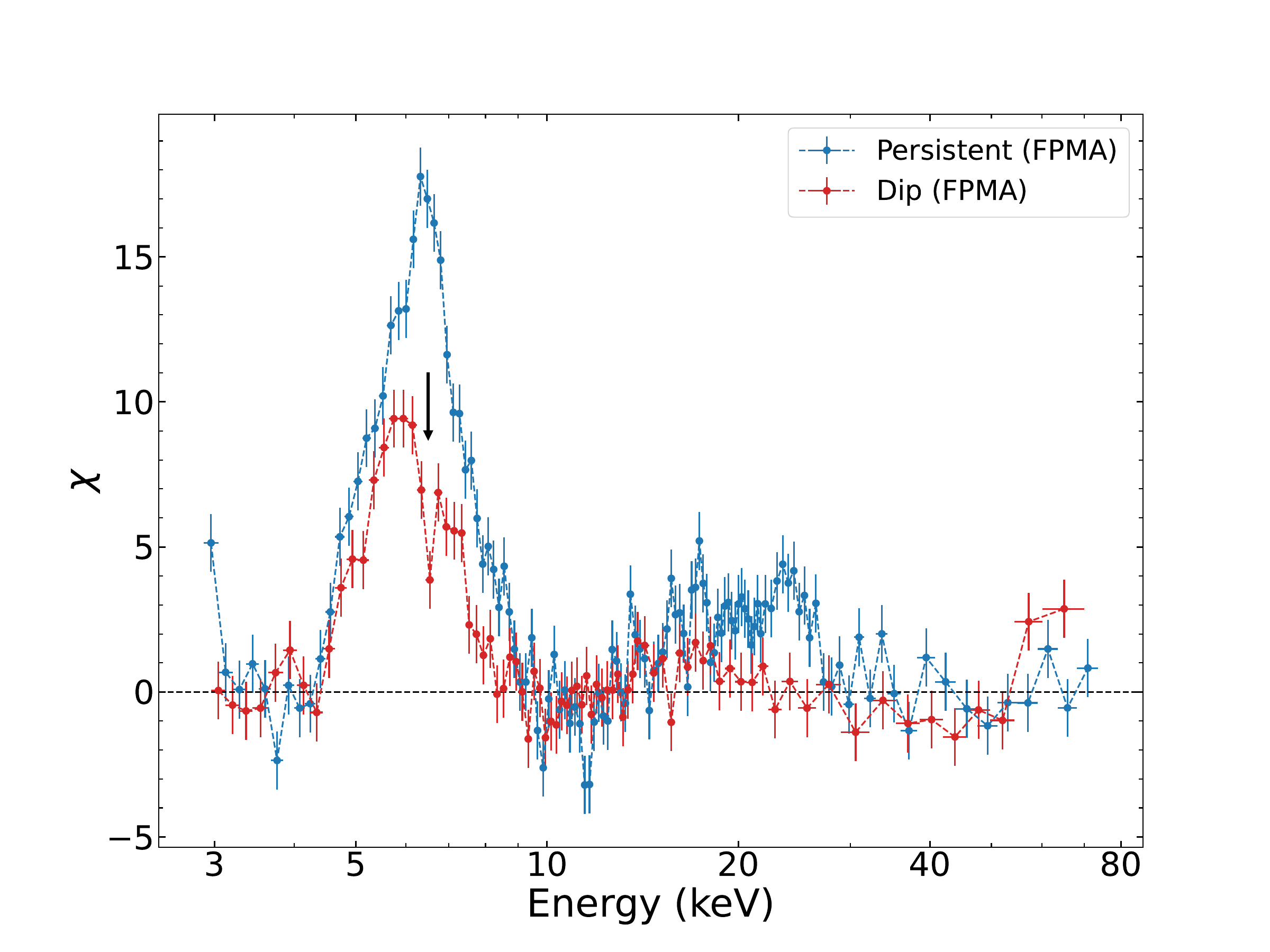}
\includegraphics[width=.50\textwidth, angle=0, trim={2cm 1cm 3cm 1cm}]{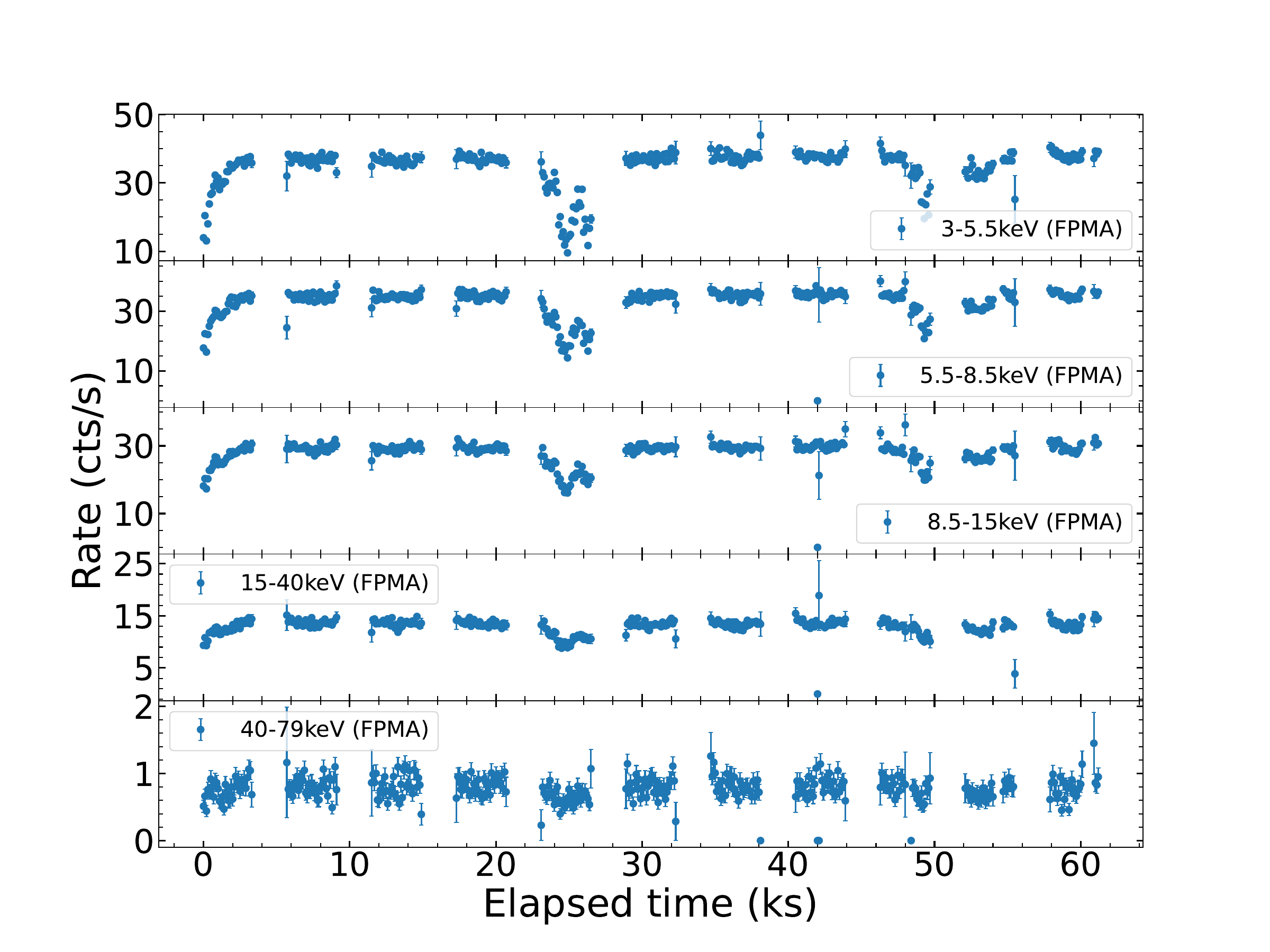}
\caption{Left: Residuals for Epoch~1 persistent and dip FPMA spectra based on an absorbed \texttt{diskbb+cutoffpl} fit to the \textit{NuSTAR} data in the energy range $3-4.5\,\mathrm{keV}$, $8.5-15\,\mathrm{keV}$ and $40-79\,\mathrm{keV}$, extrapolated over the complete energy band. The black arrow indicates a prominent absorption dip super-imposed on the broad Fe K$\alpha$ line of the dip spectra. The spectra have been rebinned for plotting purposes. Right: \textit{NuSTAR} FPMA light curves in the $3-5.5\,\mathrm{keV}$, $5.5-8.5\,\mathrm{keV}$, $8.5-15\,\mathrm{keV}$, $15-40\,\mathrm{keV}$ and $40-79\,\mathrm{keV}$ bands, from top to bottom panels respectively.}
\label{fig:mo_phen_dip-pers_lc}
\end{figure*}

We fit the absorbed \texttt{diskbb+cutoffpl} model to the dip and the persistent spectra separately following the approach described above. Although the dip spectra do not require a disk component, it was included for consistency. Both spectra reveal strong reflection features with an obvious absorption line super-imposed on the the broad Fe K$\alpha$ line in the dip spectra as shown in the left panel of Fig. \ref{fig:mo_phen_dip-pers_lc}.
The figure also reveals that the width of the broad Fe K$\alpha$ line from the dip spectra is significantly broadened, almost comparable to that from the persistent spectra. This indicates that even though most of the relativistically-broadened reflected spectra are produced in the inner disk close to the black hole, during the dip intervals when the absorber passes the line of sight, it obstructs mostly the soft X-ray photons while the harder photons get through as evident from the right panel of Fig. \ref{fig:mo_phen_dip-pers_lc}, where at harder X-ray energies, dipping tends to become less prominent. 
This is further supported by the spectral hardening at low count rates observed during this epoch, shown in Fig. \ref{fig:hid_nu}. It is also likely that the absorber is a partial covering absorber, in which case some flux may leak through unabsorbed during dips and could contribute to the broadness of its Fe K$\alpha$ line.


\begin{figure*}
\includegraphics[width=.50\textwidth, angle=0, trim={3cm 1cm 2cm 1cm}]{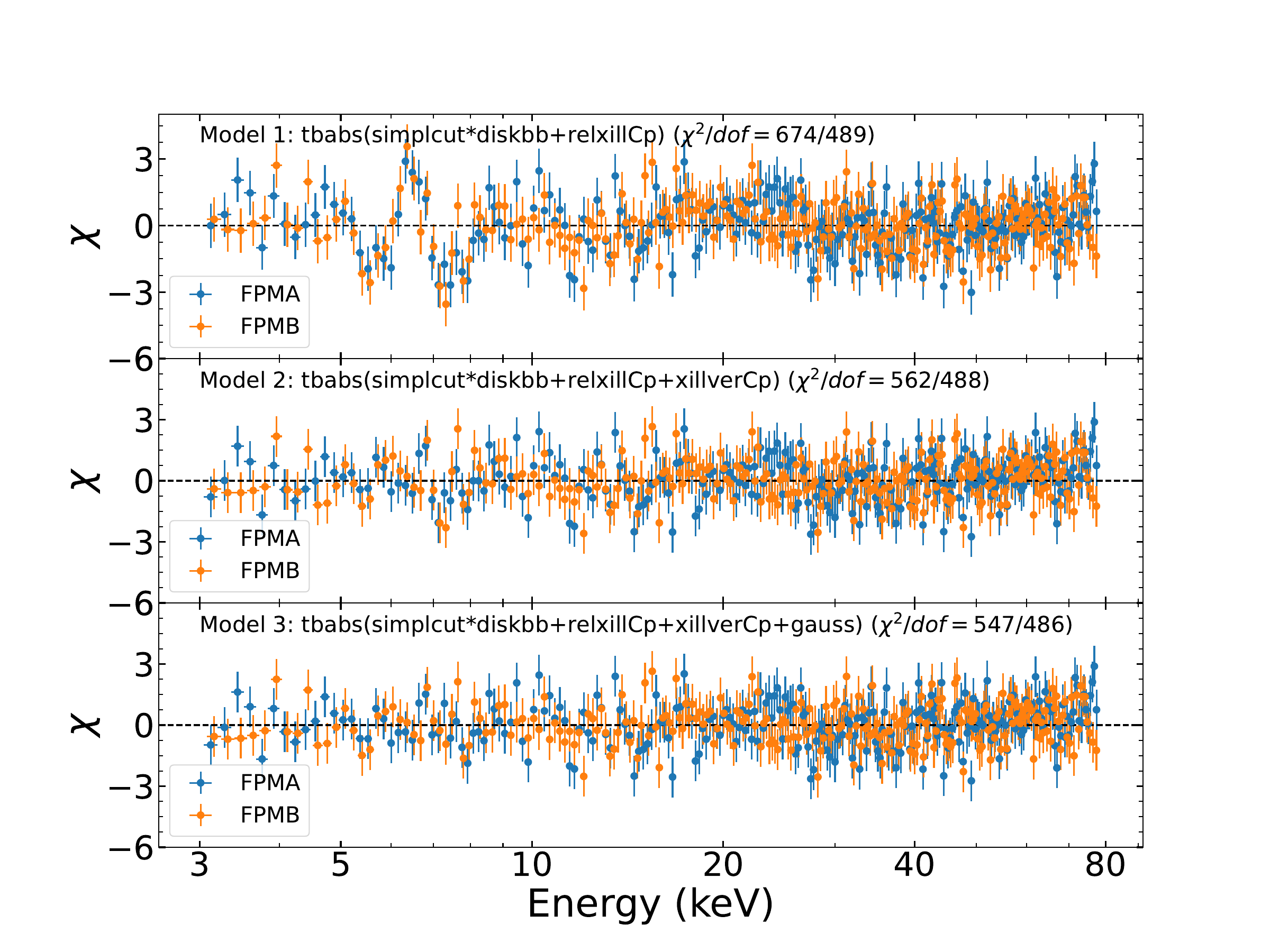}
\includegraphics[width=.50\textwidth, angle=0, trim={2cm 1cm 3cm 1cm}]{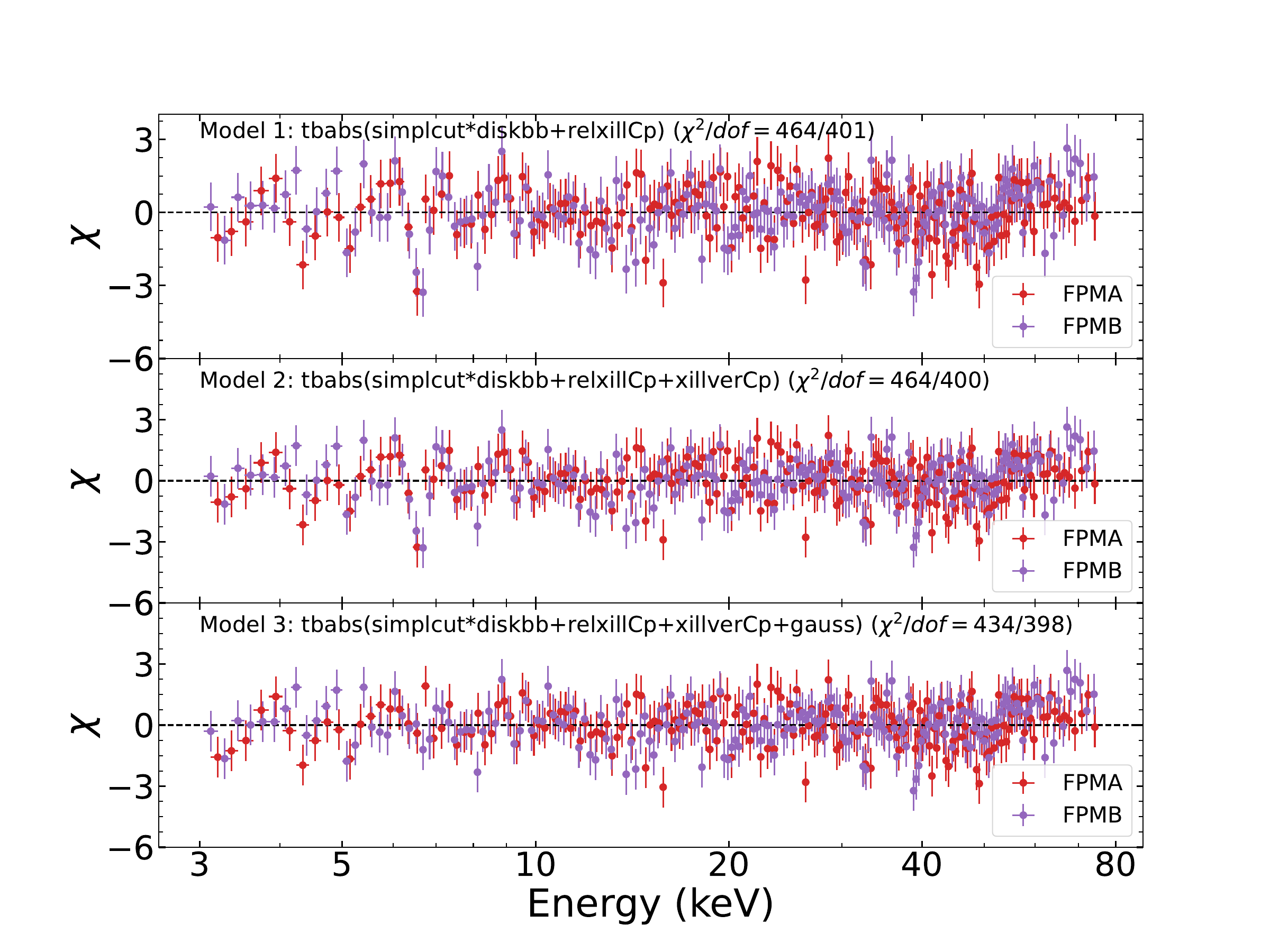}
\caption{Left: Best-fit residuals from model fits to the \textit{NuSTAR} FPMA/B persistent spectra. Right: Best-fit residuals from model fits to the \textit{NuSTAR} FPMA/B dip spectra.}
\label{fig:spec1_p-d}
\end{figure*}

To model the reflection features in the persistent spectra of Epoch 1, we started by fitting the model \texttt{cons*tbabs*(simplcut*diskbb+relxillCp)} to the data. \texttt{simplcut} is an extension of \texttt{simpl}, an empirical Comptonization model that self-consistently scatters a fraction of disk seed photons into a power law \citep{2009PASP..121.1279S}. This model gave an unacceptable fit with $\chi^2/dof=674/489$. The residual plot of Fig. \ref{fig:spec1_p-d} (left) shows that while the model reproduces the relativistic reflection features, 
there is the presence of the narrow component of the Fe K$\alpha$ emission line as well as a possible absorption feature around $7\,\mathrm{keV}$ that is not accounted for. Narrow Fe K${\alpha}$ line has been seen in the \textit{NuSTAR} spectra of a number of BHXBs but its origin is still unknown \citep[see e.g.,][]{2016ApJ...826...87W, 2017ApJ...839..110W, 2018ApJ...855....3T, 2018ApJ...852L..34X, 2018ApJ...865...18X}. Possible origins could be line emission from the stellar wind of the donor star or distant reflection by a flared disk. On the other hand, when accompanied by absorption features, the line complex may originate from accretion disk wind close to the plane of the disk. 

To account for possible contributions from distant reflection to the observed narrow iron line, we included the unblurred reflection model \texttt{xillverCp} \citep{2010ApJ...718..695G} for which we set the ionization parameter log~[$\xi/\mathrm{erg\,cm\,s^{-1}}]=0$ assuming the reflecting material is nearly neutral, and its reflection fraction R$_{f}=-1$ --- as with \texttt{relxillCp} --- so that the model only provides the reflection spectrum and not the continuum (as we assume that the distant reflection is produced by the same illuminating continuum responsible for the relativistic reflection signal). Besides the normalization, all other parameters of the model are tied to those of \texttt{relxillCp}. This improved the fit considerably, with $\Delta\chi^{2}=112$ for one additional free parameter, giving $\chi^2/dof=562/488$. Finally, to fit for the possible absorption feature around $7\,\mathrm{keV}$, we included the Gaussian line model \texttt{gauss} allowing for only negative normalization, with the width $\sigma$ frozen at $10\,\mathrm{eV}$. This provided a slightly improved fit with $\Delta\chi^{2}=15$ for 2 additional free parameters, giving $\chi^{2}/dof=547/486$.
The best-fit line energy is 
$E_{\rm{abs}}=7.23\pm{0.08}\,\mathrm{keV}$, close to the Fe K absorption edge at $7.112\,\mathrm{keV}$. Using the \texttt{edge} model in place of \texttt{gauss} also reproduce the feature, giving $\chi^{2}/dof=551/486$, with the edge energy $E_{\rm{edge}}=7.0\pm{0.1}\,\mathrm{keV}$. 

Although the dip spectra are insensitive to the inclusion or not of a putative disk contribution, for consistency, we started by fitting the model \texttt{cons*tbabs(simplcut*diskbb+relxillCp)} to the data, with $R_{in}$ set to ISCO and $a_*$ left free to vary as with the persistent spectra. This gives a fairly acceptable fit with $\chi^{2}/dof=464/401$. 
We then include \texttt{xillverCp} to model possible contributions from distant reflection as was done for the persistent spectra. This has a negligible effect on the fit parameters, giving $\chi^{2}/dof=464/400$. The residual plot of Fig. \ref{fig:spec1_p-d} (right) shows a strong absorption line feature around $6.6\,\mathrm{keV}$ which necessitated the inclusion of the negative-normalization Gaussian line model. Here, the line width is fixed at $50\,\mathrm{eV}$ since the line is noticeably broadened. This improved the fit significantly, with $\Delta\chi^{2}=30$ for 2 additional free parameters, giving $\chi^{2}/dof=434/398$. The line is centered at $E_{\rm{abs}}=6.59\pm{0.05}\,\mathrm{keV}$ -- close to the energy of the He-like {Fe~\sc{xxv}} absorption line. 

In an attempt to better understand the nature of the dip spectra, we jointly fitted both the dip and the persistent spectra with the parameters of the dip spectra tied to the best-fit parameters of the persistent spectra except for the Gaussian absorption lines as well as the normalizations of \texttt{relxillCp} and \texttt{xillverCp}. This yielded a barely acceptable fit with $\chi^{2}/dof = 1303/895$ --- confirming that indeed the dip spectra have been mostly cut off at low energies, significantly affected by the obscuring material along the line of sight. To probe this, we included an extra absorption column modeled with \texttt{tbabs} for which we set the column density to zero for the persistent spectra and allowed it to be free for the dip spectra. This improved the fit significantly with $\Delta\chi^{2}=272$ for one additional free parameter, giving $\chi^{2}/dof = 1031/894$. The column density of the second absorber is $N_{\rm{H}}=1.8\pm{0.2}\times10^{22}\mathrm{cm^{-2}}$, about a factor of six higher than the line of sight neutral column density in the direction to the source. This strongly supports the position that the principal difference between the persistent and the dip spectra is that the dip spectra corresponds to the persistent spectra but seen through an absorbing material.

\begin{figure*}
\includegraphics[width=.5\textwidth, angle=0, trim={3cm 1cm 1cm 1cm}]{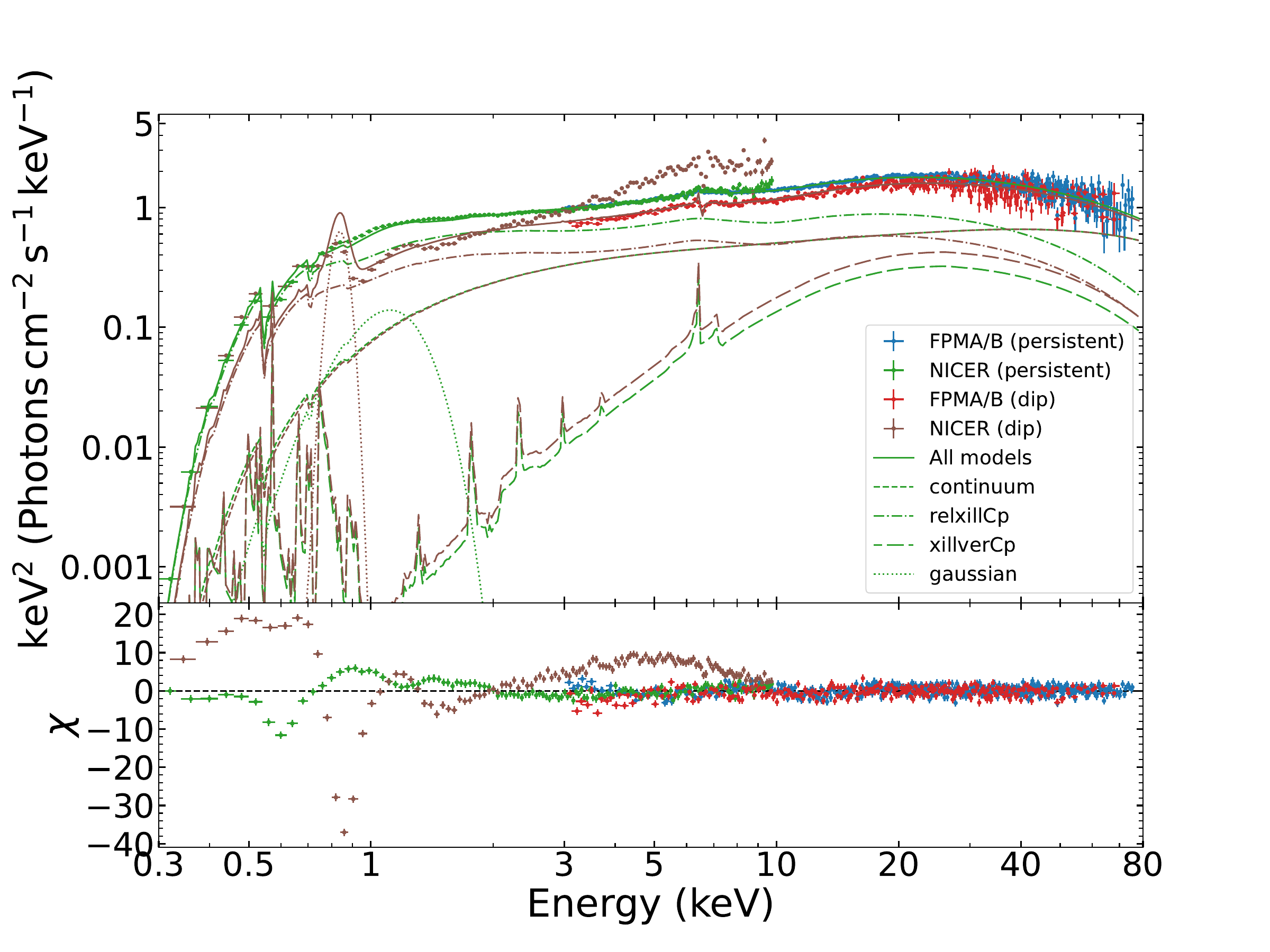}
\includegraphics[width=.5\textwidth, angle=0, trim={2cm 1cm 2cm 1cm}]{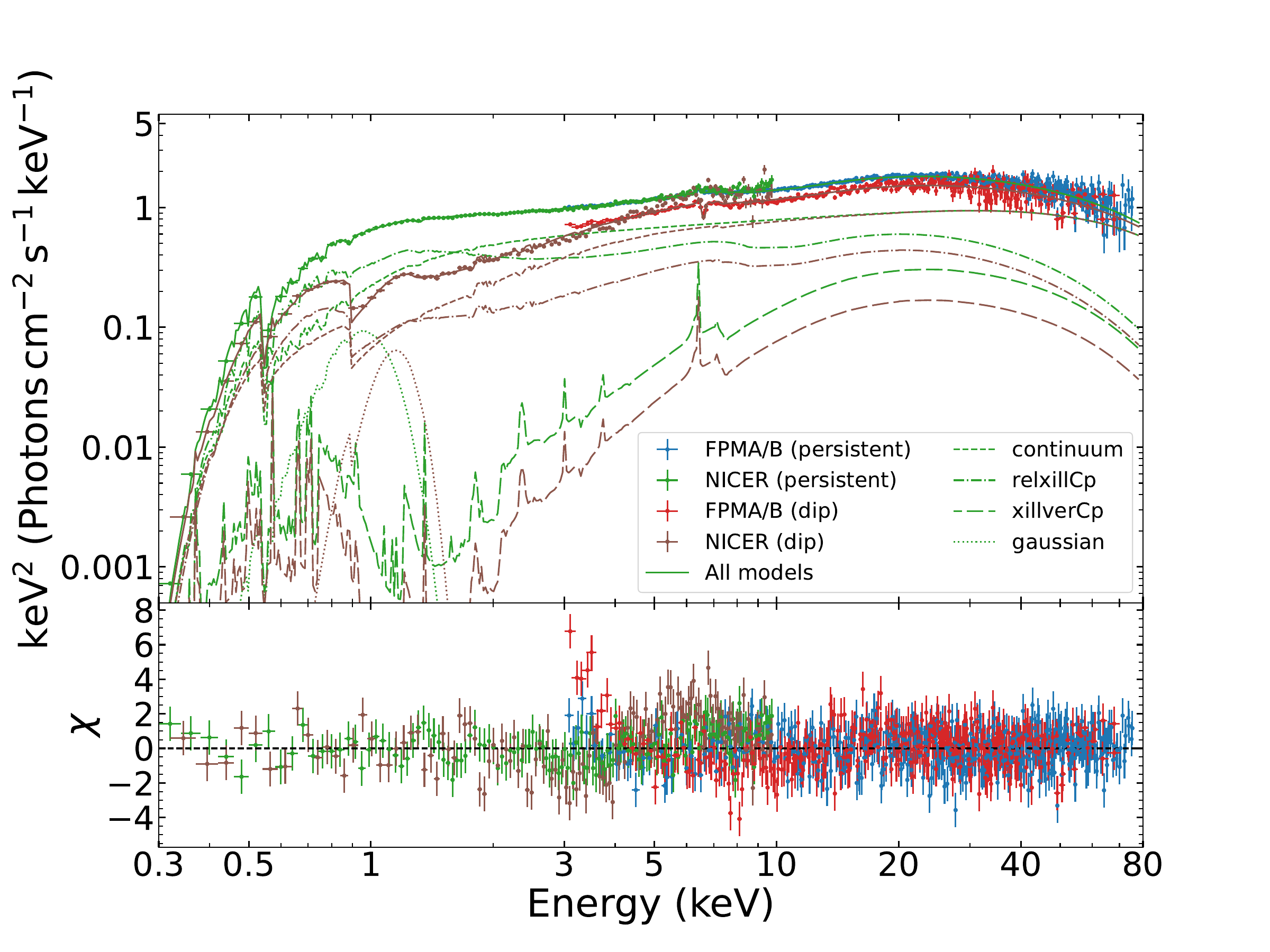}
\caption{Left: Model fit to the \textit{NICER}+\textit{NuSTAR} data of Epoch 1, where an additional \texttt{tbabs} accounts for the extra column of absorption linked to the obscurer responsible for the light-curve dips. Right: \texttt{XSTAR} table model grid is used instead of the additional \texttt{tbabs}. For both plots, the data and the different model components are shown as described by the labels. For each model component, one curve fits for the persistent spectra (green) and one for the dip spectra (brown).}
\label{fig:spec_nustar+nicer_1}
\end{figure*}

\subsubsection{\textit{NuSTAR}+\textit{NICER}} \label{subsubsec:three-one-two}
Since \textit{NICER} has a larger collecting area than \textit{NuSTAR}, greater energy resolution and extends down to $\sim0.3\,\mathrm{keV}$, it is able to better constrain the column density and provide more insight into the nature of the absorber. Therefore, we extract \textit{NICER} spectra corresponding to the dip and the persistent intervals like we did for \textit{NuSTAR}. In doing this, we only considered times for which both observations are strictly simultaneous. This corresponds to the shaded region in the upper left sub-panels within Fig. \ref{fig:lc_nustar-nicer}. During fitting, the parameters of the \textit{NICER} spectra are tied to those of the corresponding \textit{NuSTAR} spectra from the best fit described in \S~\ref{subsubsec:three-one-one} above. Additional features, present in the \textit{NICER} data between $\sim0.3\,\mathrm{keV}$ and $\sim1\,\mathrm{keV}$, are modeled with \texttt{gauss} and \texttt{edge} (see Table \ref{tab:mo_table}). It is unclear whether these residuals are strictly astrophysical or instrumental (i.e. relating to calibration) in nature. This is partly because absorption edges structures and related effects are more complicated than captured by \texttt{tbabs} and \texttt{tbfeo} as well as most other interstellar medium (ISM) absorption models. The model provided an unacceptable fit with $\chi^{2}/dof = 11,117/998$, suggesting that the additional $N_{\rm{H}}$ alone from \texttt{tbabs} that fits for the \textit{NuSTAR} dip spectra is not sufficient to reproduce the dips when \textit{NICER} spectra are considered, as shown in the left panel of Fig. \ref{fig:spec_nustar+nicer_1}. 

Proceeding, we generated a table model in \texttt{XSTAR} \citep{2001ApJS..133..221K} to fit for the absorption features. This has the added advantage of being able use an input continuum spectrum tailored to our data. To do this, we customize the \texttt{XSTAR} photoionization grid for the hard state spectrum of \rm{MAXI~J1803-298} 
using an input spectral file generated from fitting a simple disk blackbody plus cutoff power-law model to the \textit{NICER} data of the dip spectra.
The grid covers the parameter space of $10^{18}\,\mathrm{cm^{-2}}\leq N_{\rm{H}}\leq 10^{24}\,\mathrm{cm^{-2}}$ and $0\leq \mathrm{log}~[\xi^{xstar}/\mathrm{erg\,cm\,s^{-1}}] \leq 5$. For the computation, we assumed a source luminosity of $10^{38}\,\mathrm{erg\,s^{-1}}$ 
and gas density of $10^{14}\mathrm{cm^{-3}}$. 
The full model is now \texttt{cons*tbfeo*XSTAR*(simplcut*diskbb+relxillCp+ xillverCp+gauss+gauss)*edge}. During fitting, 
we allowed $N_{\rm{H}}^{xstar}$, the column density from \texttt{XSTAR}, and the ionization parameter ${\rm{log}}[\,\xi^{xstar}/\mathrm{erg\,cm\,s^{-1}}]$, to be free between the persistent and the dip spectra. We linked the velocity shift between the persistent and the dip spectra and set it to zero.
The best fit with this model gives $\chi^{2}/dof=1552/982$, 
shown in Fig. \ref{fig:spec_nustar+nicer_1} (right). The cross-calibration constant of the \textit{NICER}-dip spectra is however fairly low, at $0.61\pm{0.01}$, relative to the \textit{NuSTAR}-dip spectra. Also, slight discrepancy is still evident between the individual \textit{NICER} and their corresponding \textit{NuSTAR} spectra. 
The best-fit $N_{\rm{H}}^{xstar}$ for the dip and the persistent spectra are $1.29^{+0.20}_{-0.10}\times10^{23}\,\mathrm{cm^{-2}}$ and $4\pm{1}\times10^{21}\,\mathrm{cm^{-2}}$, respectively. The ionization parameter for the dip spectra is ${\rm{log}}[\,\xi^{xstar}/\mathrm{erg\,cm\,s^{-1}}]=1.46\pm{0.02}$ while for the persistent spectra, it is  ${\rm{log}}[\,\xi^{xstar}/\mathrm{erg\,cm\,s^{-1}}]=0.88^{+0.04}_{-0.06}$. 
Fixing $N_{\rm{H}}^{xstar}$ for the persistent spectra to the lowest allowed value of $10^{18}\,\mathrm{cm^{-2}}$ worsens the fit considerably. This indicates that some level of absorption, possibly from outflowing wind or remnants from the clumps creating the dips, is also imprinted on the persistent spectra.
This best-fit model implies a near-maximum black hole spin with $a_*=0.997^{+0.001}_{-0.002}$, and a high inclination of $i=65\pm{3}\degree$. 
The best-fit photon index is $\Gamma=1.81\pm{0.01}$ with a low corresponding coronal temperature $kT_{e}=21^{+1}_{-2}\,\mathrm{keV}$. Similarly low coronal temperatures have been reported for a number of BHXBs in the bright hard state \citep{2013ApJ...775L..45M, 2015ApJ...799L...6M, 2018ApJ...852L..34X, 2018ApJ...865...18X}. 
The energies of the absorption lines in the persistent and dip spectra are $7.10^{+0.71}_{-0.95}\,\mathrm{keV}$ and $6.61\pm{0.05}\,\mathrm{keV}$, respectively. All best-fit parameters from the \textit{NuSTAR}+\textit{NICER} fit are shown in Table \ref{tab:mo_table}.

Dividing the Gaussian line normalization by the negative of its errors indicate that the $\sim6.6\,\mathrm{keV}$ absorption line present in the dip spectra has a detection significance greater than $4\sigma$. 
Extensive Monte Carlo simulations, described in Section \ref{subsec:three-five}, confirm that the $\sim6.6\,\mathrm{keV}$ line is significantly detected above the $99.9\%$ confidence interval, while the significance level of the line at $\sim7.1\,\mathrm{keV}$ in the persistent spectra is below $50\%$. 
Given that the detection significance of the $\sim7.1\,\mathrm{keV}$ line is less than $\sim1\sigma$, it is not discussed further.
For the dip spectra, if the absorption line at $6.61\pm{0.05}\,\mathrm{keV}$ is associated with the He-like {Fe~\sc{xxv}} line at $6.7\,\mathrm{keV}$, this corresponds to a velocity redshift of $4000\pm{2200}\,\mathrm{km\,s^{-1}}$. With an ionization parameter of log~$[\xi^{xstar}/\mathrm{erg\,cm\,s^{-1}}]\sim2$ for the absorber material, the line could also be linked to a less-ionized transition from Be-like {Fe~\sc{xxiii}}, possibly the relatively strong feature at $6.6288\,\mathrm{keV}$. This would correspond to a velocity redshift of $900\pm{2300}\,\mathrm{km\,s^{-1}}$. Both of these possibilities imply that the irradiated material responsible for producing the line, as well as the light-curve dips could be moving in opposite direction of typical outflowing disk winds (i.e. away from our line of sight).
This implies that the intervening material causing the dips is distinct from any disk wind material. We note that while the absolute energy calibration of \textit{NICER} allows the detection such velocities, the systematic uncertainty of \textit{NuSTAR} \citep[i.e., $40\,\mathrm{eV}$ at energies near the Fe emission features,][]{2015ApJS..220....8M}, implies shifts of this magnitude could only be marginal at best.

In an attempt to improve the fit further, we checked the role of mission-specific calibration differences in the apparent misalignment especially between \textit{NICER} and \textit{NuSTAR} dip spectra (Fig. \ref{fig:spec_nustar+nicer_1}). We replace the cross-calibration \texttt{constant} with \texttt{crabcor} \citep[see e.g.,][]{2010ApJ...718L.117S}. This provided an improved fit, with $\chi^2/dof=1200/980$. However, while most of the fit parameters are comparable to those from the preceding fit, the \texttt{crabcor} cross-calibration normalization and photon index deviation $\Delta\Gamma$ of \textit{NICER}-dip spectra relative to \textit{NuSTAR}-dip spectra are $\sim0.3$ and $\sim-0.4$, respectively. These values are likely unrealistic and may partly result from the inability of the model employed here, and the \texttt{XSTAR} grid, to capture the full complexity of the absorber properties -- thereby mimicking a change in continuum spectral shape for the \textit{NICER}-dip spectra relative to \textit{NuSTAR}-dip spectra. The shape of the dipping intervals in the lightcurves suggest that there is likely a gradient to the column density of the absorber such that the obscurer material is more densely distributed at its core than at the edges. As such, the single column density assumption employed here is an approximation at best. A detailed analysis of column density variations across the dipping intervals is beyond the scope of the present paper.

\begin{table*}[ht!]
\caption{Best-fit parameter values for \rm{MAXI~J1803-298} for all four epochs with \texttt{relxillCp} as base model.} 
\centering 
\footnotesize
\begin{tabular}{llcccccr} 
\hline\hline 
Component & parameter & Epoch 1 (P) & Epoch 1 (D) & Epoch 2 & Epoch 3 (LF) & Epoch 3 (HF) & Epoch 4\\
\hline
Gal. abs. & $N_{\rm{H}}~(10^{21}\,\mathrm{cm^{-2}})$ & 3.2(f) & 3.2(f) & 3.2(f) & 3.2(f) & 3.2(f) & 3.2(f)\\
& $O~\rm{abund}.~(solar)$ & $0.9\pm{0.1}$ & $0.9$(t) & $1$(f) & $1.31^{+0.05}_{-0.02}$ & $1.31$(t) & $1.38^{+0.06}_{-0.14}$\\
& $Fe~\rm{abund}.~(solar)$ & $0.2\pm{0.1}$ & $0.2$(t) & $1$(f) & $0.9^{+0.3}_{-0.1}$ & $0.9$(t) & $0.6^{+0.2}_{-0.2}$\\
XSTAR & $N_{\rm{H}}^{xstar}~(10^{21}\,\mathrm{cm^{-2}})$ & $4\pm{1}$ & $129^{+20}_{-10}$ & $-$ & $-$ & $-$ & $-$\\
& log~[$\xi^{xstar}/\mathrm{erg\,cm\,s^{-1}}$] & $0.88^{+0.04}_{-0.06}$ & $1.46\pm{0.02}$ & $-$ & $-$ & $-$ & $-$\\
relxillCp & $i~({\degree})$ & $65\pm{3}$ & $65$(t) & $68^{+5}_{-4}$ & $69^{+2}_{-5}$ & $69$(t) & $87^{*}$\\
& $a_*$ & $0.997^{+0.001}_{-0.002}$ & $0.997$(t) & $0.98^{+0.01}_{-0.02}$ & $0.990^{+0.002}_{-0.007}$ & $0.990$(t) & $0.998^{*}$\\
& $R_{in}~(\rm{ISCO})$ & $1.0$(f) & $1.0$(f) & $1.0$(f) & $1.0$(f) & $1.0$(f) & $1.0$(f)\\
& $R_{out}~(\rm{r_{g}})$ & $400$(f) & $400$(f) & $400$(f) & $400$(f) & $400$(f) & $400$(f)\\
& $R_{br}~(\rm{r_{g}})$ & $15.0$(f) & $15.0$(f) & $15.0$(f) & $15.0$(f) & $15.0$(f) & $15.0$(f)\\
& $q_{1}$ & $7\pm{1}$ & $7$(t) & $9^{+1}_{-2}$ & $10^{*}$ & $10$(t) & $-10^{*}$\\
& $q_{2}$ & $3.0$(f) & $3.0$(f) & $3.0$(f) & $3.0$(f) & $3.0$(f) & $3.0$(f)\\
& $\Gamma$ & $1.81\pm{0.01}$ & $1.81$(t) & $2.14^{+0.12}_{-0.04}$ & $2.113^{+0.063}_{-0.002}$ & $2.113$(t) & $2.7^{+0.2}_{-0.1}$\\
& log~[$\xi/\mathrm{erg\,cm\,s^{-1}}$] & $2.6\pm{0.1}$ & $2.6$(t) & $3.9^{+0.3}_{-0.2}$ & $3.46^{+0.02}_{-0.28}$ & $3.62^{+0.02}_{-0.25}$ & $2.9^{+0.7}_{-0.1}$\\
& log~[$N/\mathrm{cm^{-3}}]$ & $20^{*}$ & $20$(t) & $19.96^{+0.003}_{-1.000}$ & $20^{*}$ & $20$(t) & $18^{+1}_{-2}$\\
& $A_{Fe}$ (solar) & $1.5^{+0.3}_{-0.2}$ & $1.4$(t) & $5^{+5}_{-1}$ & $4.96^{+0.02}_{-1.20}$ & $4.96$(t) & $1.0^{+0.9}_{-0.3}$\\
& $kTe\, (\rm{keV})$ & $21^{+1}_{-2}$ & $21$(t) & $400^{*}$ & $53^{+44}_{-9}$ & $53$(t) & $300$(f)\\
& $Refl_{\rm{frac}}$ & $-1$(f) & $-1$(f) & $-1$(f) & $-1$(f) & $-1$(f) & $-1$(f)\\
& $norm_{\rm{relxillcp}}~(10^{-4})$ & $145^{+11}_{-18}$ & $107^{+9}_{-15}$ & $147^{+34}_{-37}$ & $129^{+6}_{-37}$ & $145^{+6}_{-42}$ & $171^{+135}_{-109}$\\
xillverCp & $\Gamma$ & $1.81$(t) & $1.81$(t) & $-$ & $-$ & $-$ & $-$\\
& $A_{Fe}$ (solar) & $1.5$(t) & $1.5$(t) & $-$ & $-$ & $-$ & $-$\\
& $kTe\,(\mathrm{keV})$ & $21$(t) & $21$(t) & $-$ & $-$ & $-$ & $-$\\
& log~[$\xi/\mathrm{erg\,cm\,s^{-1}}$] & $0$(f) & $0$(f) & $-$ & $-$ & $-$ & $-$\\
& log~[$N/\mathrm{cm^{-3}}]$ & $20$(t) & $20$(t) & $-$ & $-$ & $-$ & $-$\\
& $i\, (\rm{\degree})$ & $65$(t) & $65$(t) & $-$ & $-$ & $-$ & $-$\\
& $Refl_{\rm{frac}}$ & $-1$(f) & $-1$(f) & $-$ & $-$ & $-$ & $-$\\
& $norm_{\rm{xillvercp}}~(10^{-4})$ & $49^{+11}_{-8}$ & $27^{+7}_{-5}$ & $-$ & $-$ & $-$ & $-$\\
simplcut & $\Gamma$ & $1.81$(t) & $1.81$(t) & $2.14$(t) & $2.113$(t) & $2.113$(t) & $2.7$(t)\\
& $F_{\rm{scat}}$ & $0.43^{+0.08}_{-0.05}\dagger$ & $0.48$(t) & $0.15^{+0.10}_{-0.03}$ & $0.37^{+0.07}_{-0.01}$ & $0.33^{+0.07}_{-0.01}$ & $0.004^{+0.010}_{-0.003}$\\
& $Refl_{\rm{frac}}$ & $1$(f) & $1$(f) & $1$(f) & $1$(f) & $1$(f) & $1$(f)\\
& $kTe\, (\rm{keV})$ & $21$(t) & $21$(t) & $400$(t) & $53$(t) & $53$(t) & $300$(t)\\
diskbb & $T_{in}$ (keV) & $0.12\pm{0.01}$ & $0.12$(t) & $1.07^{+0.02}_{-0.01}$ & $0.764^{+0.019}_{-0.002}$ & $0.837^{+0.017}_{-0.003}$ & $0.840^{+0.004}_{-0.001}$\\
& norm $(10^{2})$ & $8935^{+2179}_{-1834}$ & $8935$(t) & $5.4^{+0.7}_{-0.3}$ & $12\pm{1}$ & $9\pm{1}$ & $7.1^{+0.1}_{-0.2}$\\
gauss & ${E_{\rm{abs}}}$ (keV) & $7.10^{+0.71}_{-0.95}$ & $6.61\pm{0.05}$ & $6.86^{+0.11}_{-0.08}$ & $-$ & $-$ & $6.76\pm{0.07}$\\
& $\sigma$ (keV) & $0.01$(f) & $0.05$(f) & $0.01$(f) & $-$ & $-$ & $0.05$(f)\\
& norm $(10^{-4})$ & $-1.4^{+0.6}_{-0.3}$ & $-11.2^{+1.7}_{-1.1}$ & $-3\pm{1}$ & $-$ & $-$ & $-1.9^{+0.5}_{-0.2}$\\
gauss & ${E_{\rm{abs}}}$ (keV) & $0.8\pm{0.1}$ & $1.09\pm{0.02}$ & $-$ & $-$ & $-$ & $-$\\
& $\sigma$ (keV) & $0.17^{+0.04}_{-0.03}$ & $0.13^{+0.01}_{-0.02}$ & $-$ & $-$ & $-$ & $-$\\
& norm $(10^{-2})$ & $16^{+5}_{-6}$ & $10\pm{2}$ & $-$ & $-$ & $-$ & $-$\\
edge & $E_{\rm{edge}}$ (keV) & $0.53^{+0.04}_{-0.03}$ & $0.89\pm{0.01}$ & $-$ & $0.386^{+0.005}_{-0.004}$ & $0.386$(t) & $0.38^{+0.01}_{-0.02}$\\
& $\tau_{\rm{max}}$ & $0.16^{+0.03}_{-0.06}$ & $0.8\pm{0.1}$ & $-$ & $0.6\pm{0.1}$ & $0.6$(t) & $0.6^{+0.4}_{-0.3}$\\
$\chi^2/dof$ & & \multicolumn{2}{c}{1552/982} & 480/446 & \multicolumn{2}{c}{893/1013} & 310/284\\ 
\hline
\hline 
\end{tabular}
\label{tab:mo_table} 

{Note: ``f'' implies a frozen parameter, ``t'' implies a parameter value tied to another while ``*'' indicates the parameter is pegged at its hard limit in the best fit. ``P'' and ``D'' imply persistent and dip spectra, respectively. ``LF'' and ``HF'' imply low flux and high flux spectra, respectively. For one of the $F_{scat}$ values, ``$\dagger$'' indicates that the peak value of the MCMC probability distribution ($0.48^{+0.02}_{-0.10}$) does not exactly coincide with the quoted best-fit value but they are consistent with each other within errors. 
}
\end{table*}
\subsection{Epoch 2} \label{subsec:three-two}
The second epoch \textit{NuSTAR} observation was carried out on May 13, 2021, just after the source had transitioned to the intermediate state. The accumulated spectra for FPMA and FPMB have durations of $31.6\,\mathrm{ks}$ and $32.6\,\mathrm{ks}$, respectively. Corresponding \textit{NICER} observations were not available during this epoch because \textit{NICER} was  unable to observe \rm{MAXI~J1803-298} between the 5th and the 17th of May. As evident from Fig. \ref{fig:lc_nustar-nicer}, the source exhibited significant short-term variability during this epoch, with a fractional variability amplitude $F_{var}\sim13\%$. As shown in Fig. \ref{fig:mo_refl}, the source also showed significant disk reflection features during this epoch. Model fits to this observation are extensively reported in \citet{2023ApJ...949...70C} and so are not repeated here, the analysis is briefly discussed for completeness. 

\begin{figure*}
\includegraphics[width=.5\textwidth, angle=0, trim={3cm 1cm 1cm 1cm}]{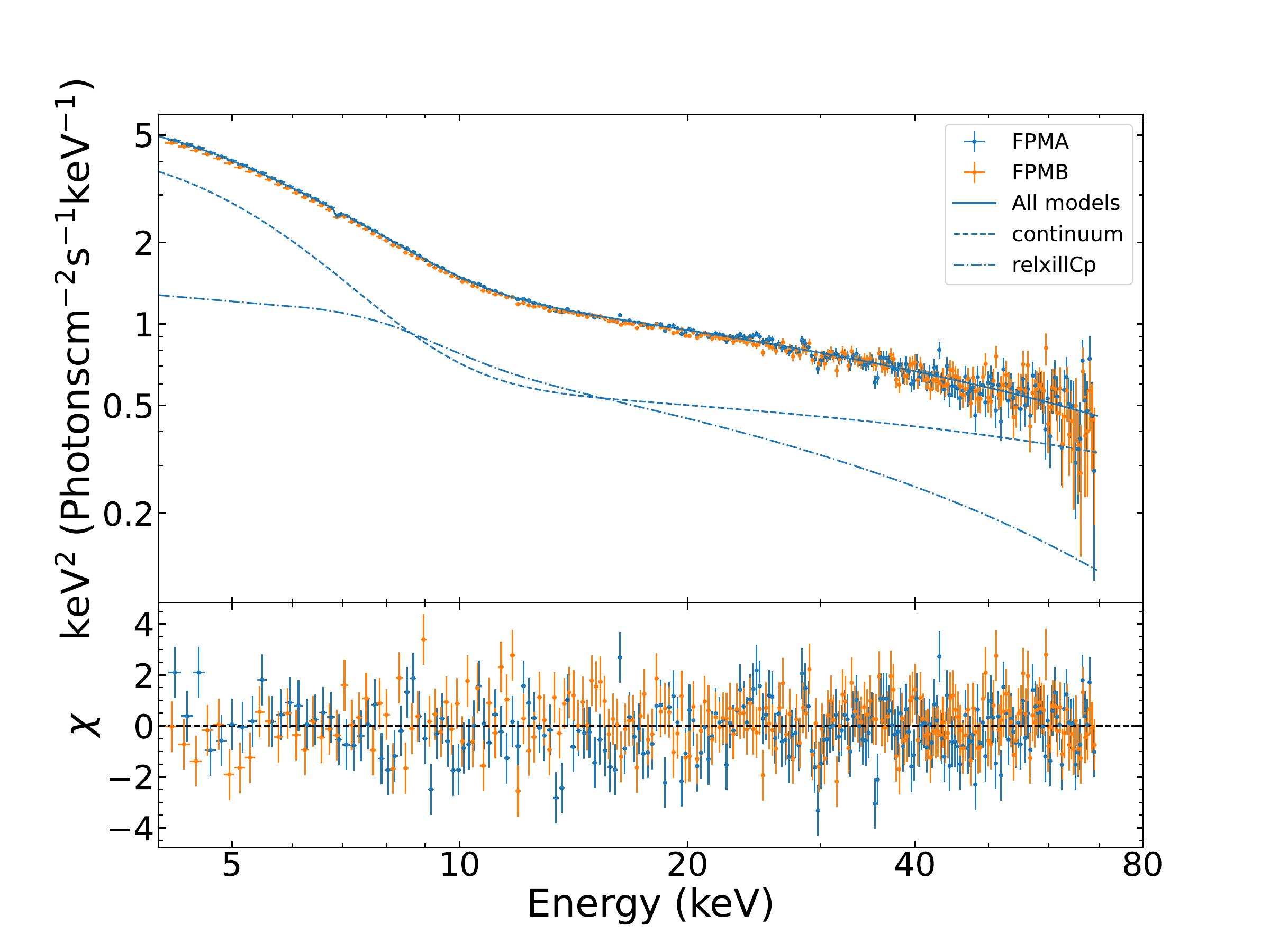}
\includegraphics[width=.5\textwidth, angle=0, trim={1cm 1cm 3cm 1cm}]{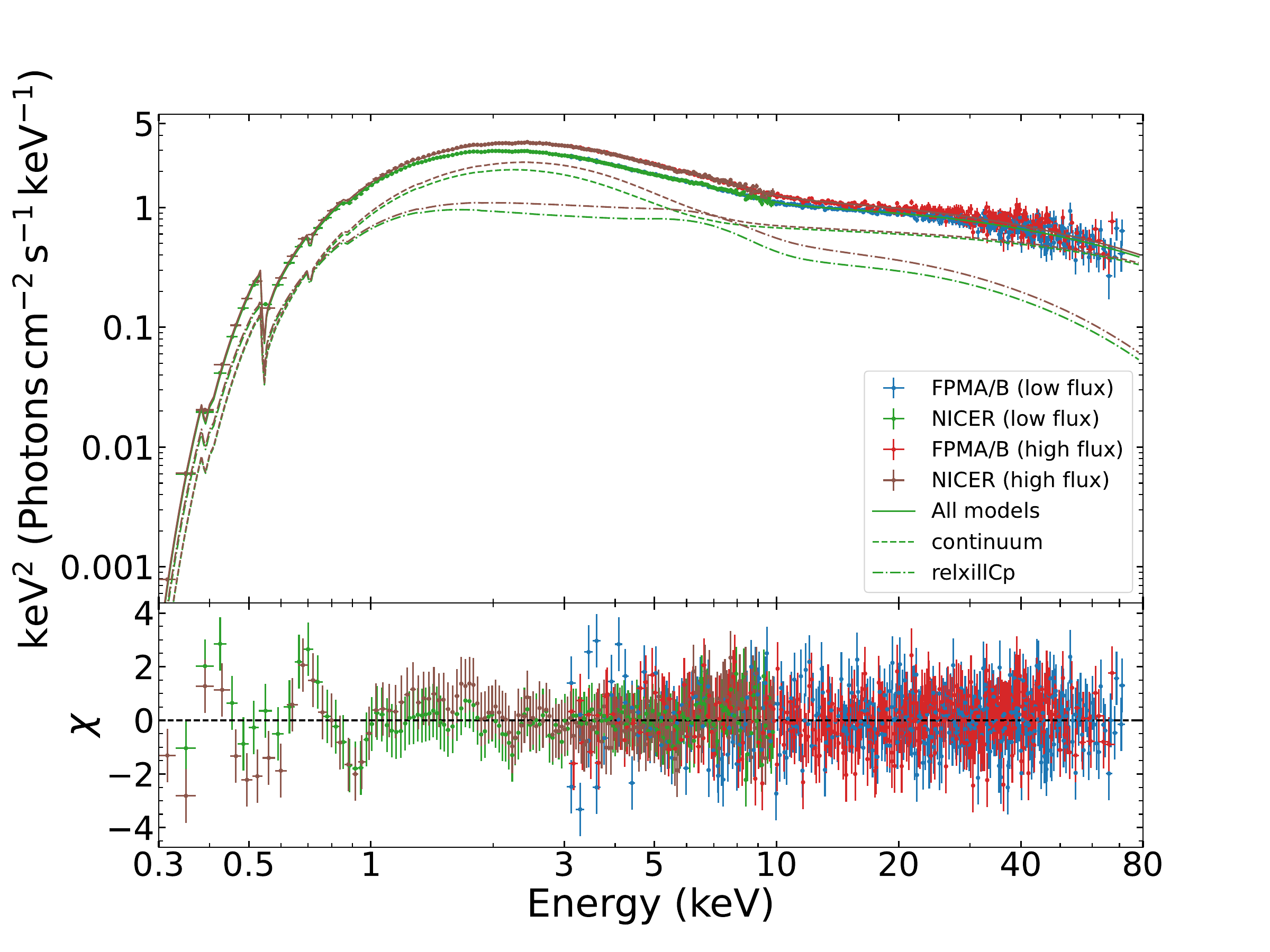}
\caption{Left: Best-fit model to the \textit{NuSTAR} FPMA/B spectra of Epoch 2 with the model components shown as described in the plot label. Right: Best-fit model to the \textit{NICER}+\textit{NuSTAR} FPMA/B spectra of Epoch 3. The model components are shown as described in the plot label, where, for each of the components, one curve fits for the low flux spectra (green) and one for the high flux spectra (brown).}
\label{fig:epoch_2-3_mod}
\end{figure*}

The best-fit model to Epoch 2 is \texttt{cons*tbfeo* (simplcut*diskbb+relxillCp+gauss)} with the best-fit parameters shown in Table \ref{tab:mo_table} and the spectral plot shown in Fig. \ref{fig:epoch_2-3_mod} (left). The \texttt{gauss} model fits for the absorption line prominently detected during this epoch with the width fixed to $10\,\mathrm{eV}$. The best-fit line energy is $E_{\rm{abs}}=6.86^{+0.11}_{-0.08}\,\mathrm{keV}$ with an estimated detection significance of $\sim3\sigma$ --- from dividing the \texttt{gauss} normalization by its negative error. Monte Carlo simulations confirm the line detection significance to be $99.9\%$. If the line is associated with the He-like {Fe~\sc{xxv}} absorption line at $6.7\,\mathrm{keV}$, then the wind material responsible has an outflow velocity of $7200^{+4900}_{-3600}\,\mathrm{km\,s^{-1}}$. If associated with the more highly ionized {Fe~\sc{xxvi}} (at $6.97\,\mathrm{keV}$) however, then the intervening material is inflowing at a velocity of $4700^{+4700}_{-3400}\,\mathrm{km\,s^{-1}}$. However, the latter velocity shift may also be consistent with zero due to \textit{NuSTAR's} energy calibration uncertainty at these energies.

\subsection{Epoch 3: High vs. Low Flux Spectra} \label{subsec:three-three}
The \textit{NuSTAR} Epoch 3 observation of \rm{MAXI~J1803-298} was carried out while the source was still in the intermediate state. The accumulated FPMA and FPMB integration times are $12.9\,\mathrm{ks}$ and $13.1\,\mathrm{ks}$, respectively. 
As shown in Fig. \ref{fig:lc_nustar-nicer}, the Epoch 3 observation shows interesting features in the light curves, including dips and, more importantly, intervals of relatively low and subsequently high flux (shaded regions in the lower left sub-panels of Fig. \ref{fig:lc_nustar-nicer}). The timing of the flux variability is aligned between \textit{NuSTAR} and \textit{NICER}, with no noticeable energy-dependent time delays.

To probe the origin of the flux variability, we extracted spectra separately for the low and the high flux intervals by generating GTIs corresponding to the respective time periods. We call these the ``low flux'' (LF) and the ``high flux'' (HF) spectra as depicted in Fig. \ref{fig:lc_nustar-nicer}. For spectral modeling, we started by fitting jointly to both \textit{NuSTAR} spectra (i.e. the low and the high flux spectra) 
the model \texttt{cons*tbabs*(simplcut*diskbb+relxillCp)} over the complete energy band $3-79\,\mathrm{keV}$. We tied all the parameters of the high flux spectra to those of the low flux spectra except for the \texttt{relxillCp} normalization and cross-calibration constants. This provided a fit to the data with $\chi^2/dof=815/685$. The fit improved significantly when both the \texttt{diskbb} temperature and its normalization were untied, with $\Delta\chi^{2}=135$ for two additional free parameters, giving $\chi^2/dof=680/683$. The best-fit disk temperature and normalization for the low-flux spectra are $0.79^{+0.02}_{-0.01}\,\mathrm{keV}$ and $730^{+119}_{-92}$, respectively while for the high-flux spectra, they are $0.87\pm{0.02}\,\mathrm{keV}$ and $589^{+64}_{-59}$, respectively.
Untying the photon index did not improve the fit further,  
giving $\chi^{2}/dof=659/682$. The photon indices are also unchanged within errors, with values of $2.01^{+0.03}_{-0.07}$ and $2.03^{+0.04}_{-0.09}$ for the low and the high flux spectra, respectively.  This shows that the shape of the power law remains essentially the same between the low and the high flux spectra while there is a more significant change to the shape of the disk blackbody component. 

We then included the \textit{NICER} low and high-flux data to improve constraints on the lower energy part of the spectra, which appears to be driving the flux variability --- particularly \texttt{diskbb} temperature $T_{in}$ and its normalization which are most sensitive to the soft X-ray band. The spectral plot is shown in Fig. \ref {fig:epoch_2-3_mod} (right).
During the fit, all parameters of the \textit{NICER} low and high-flux data are tied to their corresponding \textit{NuSTAR} spectral values. Besides the disk blackbody temperature and its normalization, the ionization parameter, the scattering fraction
and the \texttt{relxillCp} normalization, all other parameters are left tied between the low and the high flux spectra. 
The model yielded a good fit to the data with $\chi^2/dof=893/1013$. The disk temperature and its normalization are $0.764^{+0.019}_{-0.002}\,\mathrm{keV}$ and $1150^{+101}_{-79}$, respectively for the low flux spectra and $0.837^{+0.017}_{-0.003}\,\mathrm{keV}$ and $897^{+94}_{-52}$ for the high flux spectra. The best-fit values of the ionization parameter are log~[$\xi/\mathrm{erg\,cm\,s^{-1}}]=3.46^{+0.02}_{-0.28}$ and log~[$\xi/\mathrm{erg\,cm\,s^{-1}}]=3.62^{+0.02}_{-0.25}$ for the low and the high flux spectra, respectively. 
The best-fit parameter values are reported in Table \ref{tab:mo_table}. When untied, the photon index values for the low and the high-flux spectra are again comparable within errors, giving $2.15^{+0.02}_{-0.01}$ and $2.18^{+0.03}_{-0.02}$, respectively, with $\chi^2/dof=890/1012$. 

This confirms that the difference in disk normalization between the low and the high flux spectra as well as the disk temperature is the most important driver of the flux variability. 

As a consistency check, we compared the ratio of the ionizing flux $F_{x}$ in the $1-100\,\mathrm{keV}$ range to that of the ionization parameter for the low and the high flux spectra. With the ionization parameter $\xi$ defined as;
\begin{equation}
    \xi = \frac{L}{nr^{2}} = 4\pi\frac{F_{x}}{n}
\label{eqn:one}
\end{equation}
where $L\,(=4\pi r^{2}F_{x})$ is the luminosity of the ionizing source, $n$ is the density of the irradiated material and $r$ is the distance between the source and the irradiated material. From the above relation, it follows that with all other quantities taken to be same, the ratio of the ionization parameters should be comparable to that of the ionizing fluxes. Using the simple phenomenological model \texttt{cons*tbfeo*edge(cflux*cutoffpl+diskbb+gauss)} --- where \texttt{cutoffpl}, \texttt{diskbb} and \texttt{gauss} crudely model contributions from the power law, disk blackbody and the broad iron line around $6.4\,\mathrm{keV}$ --- to fit for the low and high flux spectra separately,
the flux of the ionizing \texttt{cutoffpl} contribution in the $1-100\,\mathrm{keV}$ for the low and the high flux spectra are $8.9\pm{0.1}\times10^{-9}\,\mathrm{erg\,cm^{-2}\,s^{-1}}$ and $9.7\pm{0.1}\times10^{-9}\,\mathrm{erg\,cm^{-2}\,s^{-1}}$, respectively. The flux ratio gives $1.10\pm{0.01}$ while from Table \ref{tab:mo_table}, the ratio of their ionization parameters is $1.05^{+0.11}_{-0.41}$. These values are statistically equivalent. 


\subsection{Epoch 4: The Case for Disk Self-irradiation} \label{subsec:three-four}
The fourth \textit{NuSTAR} observation was carried out while the source was in the soft state, in the low flux regime \citep[see e.g.,][]{2022ApJ...926L..10M}. For this observation, the \textit{NuSTAR} spectra are considered only up to $25\,\mathrm{keV}$ because background counts dominate the data above this energy. As evident in Fig. \ref{fig:mo_refl}, the source also displayed significant relativistic reflection features during this observation as well as a prominent absorption line between $\sim6-7\,\mathrm{keV}$ that appears to be super-imposed on the broad Fe K$\alpha$ emission complex.

To model the \textit{NuSTAR} spectrum of the source during this epoch, we use the base model \texttt{cons*tbabs(simplcut*diskbb+relxillCp)}. This model gave an unacceptable fit to the \textit{NuSTAR} data with $\chi^{2}/dof = 210/141$. 
The residual plot 
reveals that the model could not account for the prominent absorption feature between $6-7\,\mathrm{keV}$. To remedy this, we included \texttt{gauss}, restricting the normalization to negative values and fixing the width at $\sigma=50\,\mathrm{eV}$. This improved the fit considerably with $\Delta\chi^{2} = 42$ for two additional free parameters, giving $\chi^{2}/dof = 168/139$. The centroid energy of the line is $E_{\rm{abs}} = 6.75^{+0.06}_{-0.03}\,\mathrm{keV}$. 
The inclusion of the quasi-simultaneous \textit{NICER} data (observations carried out on the same day with \textit{NuSTAR} but do not exactly overlap) provided a better overall fit to the data, giving $\chi^{2}/dof=310/284$. We have also replaced \texttt{tbabs} with \texttt{tbfeo}  and included the model \texttt{edge} to mitigate the known \textit{NICER} instrumental/astrophysical features below $\sim1-2\,\mathrm{keV}$.
The model however could not constrain the spin, the inclination and the inner emissivity index, which are pegged at their hard limits of 0.998, $87\degree$ and -10 respectively.
The best-fit parameter values are shown in Table \ref{tab:mo_table} and the corresponding spectral plot is shown in Fig. \ref{fig:epoch_4_mod} (left panel).

In the $0.1-100\,\mathrm{keV}$ range, the power-law contribution to the overall flux for Epoch 4 is no more than $\sim5\%$. 
Therefore, it is unlikely that the prominent reflection features observed in this state are predominantly from the coronal X-rays shining back on the disk --- the implicit assumption in \texttt{relxillCp}. We suspect that returning disk radiation in the inner regions  due to strong GR effects within this environment may be the dominant contributor to the reflection spectrum in this state \citep[see e.g.,][]{2020ApJ...892...47C, 2021ApJ...909..146C, 2022MNRAS.514.3965D}. We therefore employed the \texttt{relxillNS} model \citep{2022ApJ...926...13G} in place of \texttt{relxillCp} for the joint \textit{NICER}+\textit{NuSTAR} data. 
The irradiating continuum of \texttt{relxillNS} depends on the temperature $T_{in}$ of \texttt{diskbb} and so we tied the blackbody temperature in \texttt{relxillNS} to the \texttt{diskbb} temperature. This is unlike in \texttt{relxillCp} where the irradiating continuum is determined by the photon index $\Gamma$ and the corona temperature $kT_{e}$. We fixed the corona temperature from \texttt{simplcut} to $kT_{e}=300\,\mathrm{keV}$ and the inner disk radius from \texttt{relxillNS} is set to $R_{ISCO}$. The model combination is \texttt{cons*tbfeo(simplcut*diskbb+ relxillNS+gauss)*edge}. 
Employing \texttt{relxillNS} improved the fit significantly with $\Delta\chi^{2}=31$, for no additional free parameter, giving $\chi^{2}/dof=279/284$. The fit parameters are also generally more physical and better constrained, e.g., the spin parameter, inclination and the inner emissivity index have values $a_*=0.992^{+0.004}_{-0.028}$, $i=71^{+7}_{-8}$ and $q_{1}=6^{+4}_{-3}$, respectively. The complete list of best-fit parameters is presented in Table \ref{tab:relxillns} and the associated spectral plot is shown in the right panel of Fig. \ref{fig:epoch_4_mod}. 

With this fit, the energy of the iron absorption line is consistent at $E_{\rm{abs}}=6.74^{+0.05}_{-0.06}\,\mathrm{keV}$, and has a detection significance greater than $3\sigma$ --- obtained by dividing the line normalization by its negative error. Through Monte Carlo simulations (see Section \ref{subsec:three-five}), we estimate the line detection significance to be greater than $99.9\%$. If this line is associated with the He-like {Fe~\sc{xxv}}, it corresponds to a blueshift of $1800^{+2200}_{-2700}\,\mathrm{km\,s^{-1}}$, potentially indicative of a disk wind. Again the \textit{NuSTAR} energy calibration uncertainties mean that the velocity may be consistent with zero. 
Better constraints on the velocity shift and the outflowing material properties can be probed by employing photo-ionization models like \texttt{XSTAR} and \texttt{SPEX} \citep{1996uxsa.conf..411K} with data from higher resolution instruments. 
We note that while the power law contribution over the $0.1-100\,\mathrm{keV}$ band is negligible, it tends to dominate above $8.8\,\mathrm{keV}$ -- the relevant energy for {Fe~\sc{xxv}} ionization. This suggests that the non-thermal continuum flux plays a significant role in producing the absorption line through interaction with the wind material.
\begin{table}
\caption{The best-fit parameter values for Epoch 4 of \rm{MAXI~J1803-298} with \texttt{relxillNS} in place of \texttt{relxillCp}.} 
\centering 
\small
\begin{tabular}{llrccccc} 
\hline\hline 
Component & parameter & Epoch 4\\
\hline
Gal. abs. & $N_{\rm{H}}~(10^{21}\,\rm{cm^{-2}})$ & 3.2(f)\\
& $O~\rm{abund}.$ (solar) & $1.2\pm{0.1}$\\
& $Fe~\rm{abund}.$ (solar) & $0.8^{+0.2}_{-0.3}$\\
relxillNS & $i\, (\rm{\degree})$ & $71^{+7}_{-8}$\\
& $a_*$ & $0.992^{+0.004}_{-0.028}$\\
& $R_{in}~(\rm{ISCO})$ & $1.0$(f)\\
& $R_{out}~(\rm{r_{g}})$ & $400$(f)\\
& $R_{br}~(\rm{r_{g}})$ & $15.0$(f)\\
& $q_{1}$ & $6^{+4}_{-3}$\\
& $q_{2}$ & $3.0$(f)\\
& log~[$\xi/\mathrm{erg\,cm\,s^{-1}}$] & $3.0^{+0.5}_{-0.3}$\\
& log~[$N/\mathrm{cm^{-3}}]$ & $19^{*}$\\
& $A_{Fe}$ (solar) & $4^{+3}_{-1}$\\
& $kT_{bb}\, (\rm{keV})$ & $0.79$(t)\\
& $Refl_{frac}$ & $-1$(f)\\
& $norm_{\rm{relxillNS}}~(10^{-4})$ & $35^{+26}_{-17}$\\
simplcut & $\Gamma$ & $2.6^{+0.2}_{-0.1}$\\
& $F_{\rm{scat}}$ & $0.02\pm{0.01}$\\
& $Refl_{frac}$ & $1$(f)\\
& $kTe\, (\rm{keV})$ & $300$(f)\\
diskbb & $T_{in}$ (keV) & $0.79^{+0.02}_{-0.03}$\\
& norm $(10^{2})$ & $8.4^{+0.5}_{-0.3}$\\
gauss & ${E_{\rm{abs}}}$ (keV) & $6.74^{+0.05}_{-0.06}$\\
& $\sigma$ (keV) & $0.05$(f)\\
& norm $(10^{-4})$ & $-1.8^{+0.3}_{-0.5}$\\
edge & $E_{\rm{edge}}$ (keV) & $0.41^{+0.05}_{-0.03}$\\
& $\tau_{\rm{max}}$ & $0.21^{+0.07}_{-0.12}$\\ 
$\chi^2/dof$ & & 279/284\\
\hline
\hline 
\end{tabular}
\label{tab:relxillns} 

{Note: ``f'' implies a frozen parameter, ``t'' implies a parameter value tied to another and ``*'' indicates the parameter is pegged at its hard limit in the best fit.}
\end{table}

\subsection{Joint Spectral Fit} \label{subsec:three-five}
Recent results from reflection spectroscopy of BHXBs are revealing that important spectral parameters, including spin and inclination, are best constrained using data from multiple observations \citep[see e.g.,][]{2015ApJ...813...84G, 2021ApJ...909..146C, 2023ApJ...946...19D}. Therefore, to obtain joint constraints on the spin and the inclination of \rm{MAXI~J1803-298}, we jointly fitted \textit{NICER} and \textit{NuSTAR} data from all four epochs using a model combination made up of the best-fit model from each individual epoch i.e \texttt{cons*tbfeo*XSTAR*(simplcut*diskbb+relxillCp+ relxillNS+xillverCp+gauss+gauss)*edge}. The spin, inclination and the iron abundance are tied across all data groups. Other model parameters for each data groups were initially fixed to their best-fit values as obtained from fits to the individual spectra (shown in Tables \ref{tab:mo_table} \& \ref{tab:relxillns}) and then unfrozen during subsequent fittings. 
For data from Epochs 2, 3 and 4 which were not initially fitted with \texttt{XSTAR}, $N_{H}^{xstar}$ is kept frozen at its lowest value of $10^{18}\,\mathrm{cm^{-2}}$ while log~[$\xi^{xstar}/\mathrm{erg\,cm\,s^{-1}}$] is set to zero. \texttt{xillverCp} normalization and the normalization of the second \texttt{gauss} --- meant to model the low energy feature from \textit{NICER} in Epoch 1 --- are also set to zero for Epochs 2-4. Equally, the normalization of \texttt{relxillCp} is set to zero for Epoch 4 while the normalization of \texttt{relxillNS} is set to zero for data groups from Epochs 1-3.
The model provided an acceptable fit to the overall spectra, giving $\chi^{2}/dof = 3252/2735$. The best-fit parameters for each of the epochs are mostly consistent with the values reported in Table \ref{tab:mo_table}. From the joint spectral fit, the spin and the inclination are well constrained to be $0.990\pm{0.001}$ and $70\pm{1}\,\rm{\degree}$, respectively while the iron abundance is $3.0\pm{0.2}$. 

\subsection{Absorption line Detection Significance} \label{subsec:three-six}
An F-test can over-estimate the detection significance of emission/absorption line features in a blind search as it does not take into account the possible energy range over which the line is expected nor does it account for the number of resolution elements or bins present in that energy range \citep[see e.g.,][]{2002ApJ...571..545P, 2010A&A...521A..57T}. We therefore employ Monte Carlo simulations to estimate the detection significance of the Fe K absorption lines observed in the spectra of \rm{MAXI~J1803-298} for Epochs 1, 2 and 4. To do this, we make the assumption that in the energy range $6-7.5\,\mathrm{keV}$ there is no preference to finding a line at any particular energy. We then estimate the probability distribution of detecting randomly generated lines in this energy band and compare with the observed lines \citep[see e.g.,][]{2010A&A...521A..57T, 2022ApJ...931...77D}. 

We tested the null hypothesis that a model excluding the absorption lines is adequate to reproduce the spectra as follows:
(1) Using \texttt{fakeit} in \texttt{XSPEC}, we simulated spectra from the baseline models excluding the absorption lines.
(2) The simulated spectrum is fitted with the same baseline model and the $\chi^{2}$ value stored. To minimize complications, only the energy range $5-10\,\mathrm{keV}$ is considered for this analysis.
(3) An unresolved Gaussian line, with its width frozen to the values reported in Table \ref{tab:mo_table}, is added to the baseline model. Its normalization is set to zero and allowed to vary freely between positive and negative values. We stepped the centroid energy of the line from $6\,\mathrm{keV}$ to $7.5\,\mathrm{keV}$ at intervals of $50\,\mathrm{eV}$. This blind search is meant to account for the range of energies over which we expect such a line. Each time, we make a fit and eventually store the minimum value of $\chi^{2}$. 
(4) We repeat this procedure 1000 times and generate a distribution of simulated \textit{maximum} $\Delta\chi^{2}$ values\footnote{i.e. the difference between $\chi^{2}$ from step 1 and the minimum $\chi^{2}$ values over the energy steps $6-7.5\,\mathrm{keV}$}. If the number of simulated $\Delta\chi^{2}$ values greater than or equal to the real value\footnote{the real value is the difference in $\chi^{2}$ from the baseline model fit to the data with and without the Gaussian absorption line} is $N$, then for $S$ number of simulations, the estimated significance level will be $1-N/S$ from the Monte Carlo simulation, where $N/S$ is the \textit{p-\rm{value}}.


\begin{figure*}
\includegraphics[width=.5\textwidth, angle=0, trim={3cm 1cm 1cm 1cm}]{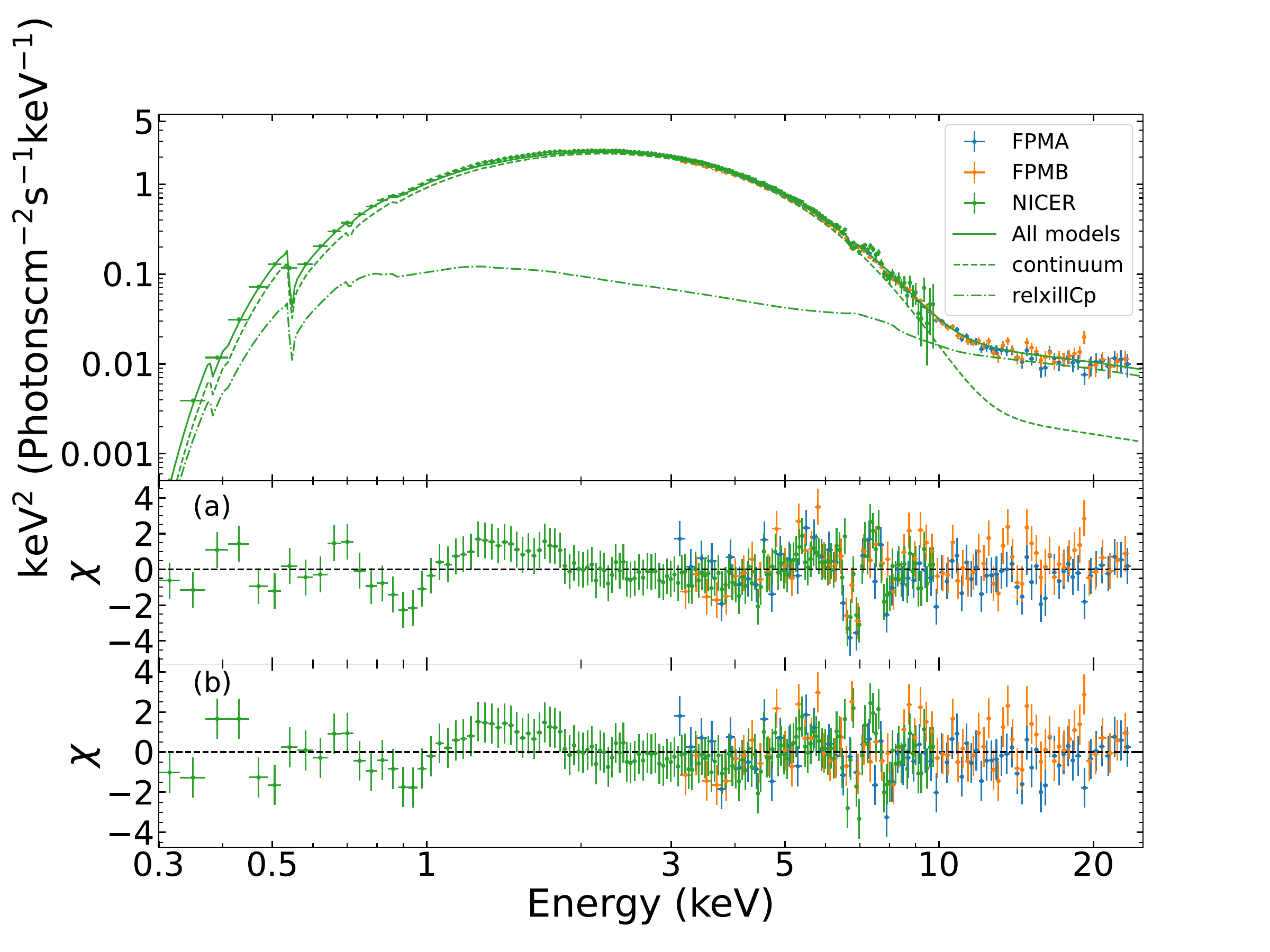}
\includegraphics[width=.5\textwidth, angle=0, trim={1cm 1cm 3cm 1cm}]{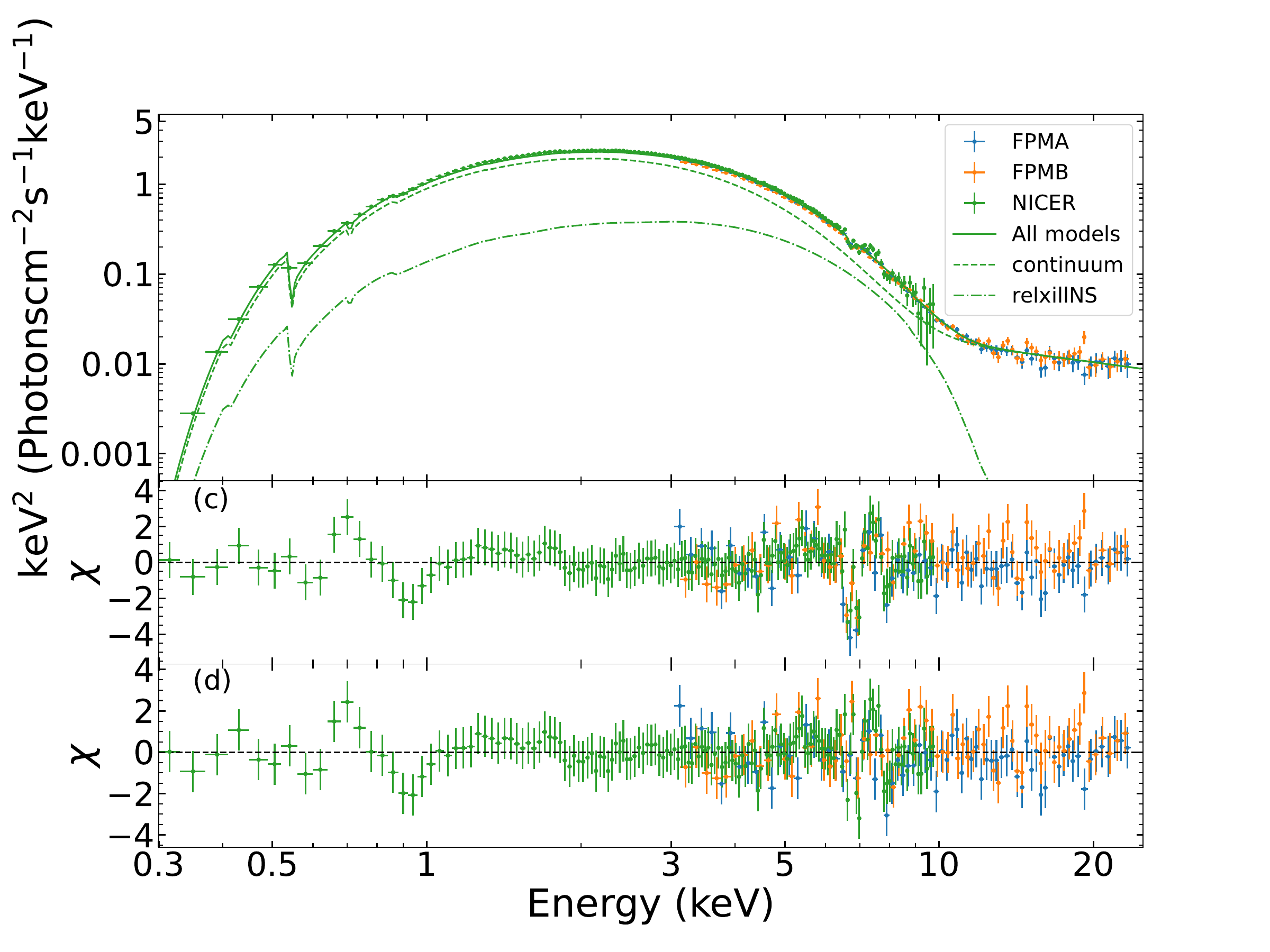}
\caption{Left: Best-fit model to the \textit{NICER}+\textit{NuSTAR} spectra of Epoch 4, where  relativistic reflection is modeled with \texttt{relxillCp}. The subplots (a) and (b) show residuals without and with the \texttt{gauss} model included, respectively, for the absorption feature at $\sim6.7\,\mathrm{keV}$. Right: Best-fit model to the \textit{NICER}+\textit{NuSTAR} spectra of Epoch 4, where relativistic reflection is modeled with  \texttt{relxillNS}. The subplots (c) and (d) show residuals without and with the \texttt{gauss} model included, respectively, for the absorption feature at $\sim6.7\,\mathrm{keV}$. For both plots, the model components are shown as indicated by the plot labels.}
\label{fig:epoch_4_mod}
\end{figure*}

\section{Discussion} \label{sec:four}
We report results from the broadband spectral analysis of the black hole candidate \rm{MAXI~J1803-298} based on data from \textit{NuSTAR} supplemented with \textit{NICER}. Using its \textit{NuSTAR} observations as baseline, we followed the spectral evolution of the source from the hard state through the intermediate state and into the soft state in order to probe and characterize its evolving spectral behavior. 
\subsection{Epoch 1: Hard State} \label{subsec:four-one}
The shape of the broadband continuum during Epoch 1 indicates the source is in the bright hard state (with $\Gamma=1.81\pm{0.01}$) --- just before transitioning to the intermediate state a few days later. 
The spectrum shows strong relativistic reflection features (see Fig. \ref{fig:mo_refl}), commonly seen in the \textit{NuSTAR} spectra of BHXBs in outburst \citep[e.g.,][]{2014ApJ...780...78T, 2015ApJ...808..122F, 2017ApJ...839..110W, 2018ApJ...852L..34X}. A narrow Fe K$\alpha$ line is also evident during this epoch, super-imposed on the broad component. We modeled this line with \texttt{xillverCp} and posit --- among other possibilities --- that the line is the product of 
distant reflection 
 \citep[see e.g.,][]{2016ApJ...826...87W, 2018ApJ...855....3T, 2018ApJ...865...18X}. 

Light-curve dips exhibited by the source during this epoch have been reported by a number of other X-ray telescopes while the source was in the hard/intermediate states \citep[e.g.,][]{2021ATel14606....1H, 2021ATel14629....1J, 2021ATel14650....1M}. 
Spectra from both the persistent and the dipping intervals show strong relativistic reflection features and a joint fit of both spectra confirms the source to be near-maximally spinning ($a_*>0.99$) and observed at a high inclination, close to edge-on ($i=65\pm{3}\degree$). This is in-line with our expectation for the source given that an inclination significantly higher than $\sim75\degree$ should give rise to periodic occultations from the companion star, which we do not see in this system. The joint fit further shows that the recurring dips can be accounted for by photo-electric absorption and Compton scattering from additional, moderately ionized material (log~[$\xi/\mathrm{erg\,cm\,s^{-1}}]\sim2$) with an absorption column of $1.29^{+0.20}_{-0.10}\times10^{23}\,\mathrm{cm^{-2}}$, compared to $4\pm{1}\times10^{21}\,\mathrm{cm^{-2}}$ during the non-dip intervals. This is about a factor of 30 higher than the non-dip intervals.
\citet{2022MNRAS.511.3922J} obtained a comparable value of $\sim2.8\times10^{23}\,\mathrm{cm^{-2}}$ for the column density and a slightly higher value of log~[$\xi/\mathrm{erg\,cm\,s^{-1}}]\sim3.7$ for the ionization parameter of the absorber based on \textit{AstroSat} data taken after the source transitioned to the intermediate state. This could potentially indicate an evolving obscurer. It should be pointed out that those authors only applied phenomenological models to fit the overall spectra, which may be adequate for their purposes considering the smaller effective area of the \textit{AstroSat} SXT instrument compared to \textit{NICER}.

Using \textit{RXTE} observations of the BHXB \rm{4U~1630-472} which also showed X-ray dips, \citet{1998ApJ...494..747T} found the column density during the dips to be about 12 times higher than during the non-dip interval.  \citet{2018ApJ...865...18X} obtained a similar conclusion in their analysis of \textit{NuSTAR} and \textit{Swift} observations of the ``dipping'' BHXB \rm{Swift~J1658.2-4242} while in the hard state. 

For Epoch 1 observation of \rm{MAXI~J1803-298}, the fact that a separate fit to the dip spectra does not require a \texttt{diskbb} component as opposed to the persistent spectra coupled with the spectral hardening at low count-rates indicates that the obscurer is close to the disk plane
and so obscures the softer disk photons more effectively. This is also supported by the energy-resolved lightcurve shown in Fig. \ref{fig:mo_phen_dip-pers_lc}, where at harder X-ray energies, the dips tend to disappear. One possible cause is a scenario where the stream of material from the companion is thicker than the scale height of the accretion disk resulting in a fraction of the stream flowing above and below the disk. According to the model by \citet{1987A&A...178..137F}, when such a material intercepts the irradiating X-ray continuum, ionization instabilities can separate the material into cold, relatively dense clouds --- which are responsible for the dips --- in a hot inter-cloud medium. The geometrical interpretation from this model however requires an extended corona. 

The observed X-ray dips have a recurrence period of $\sim7\,\mathrm{hours}$ which is most likely the orbital period of the binary system \citep[e.g.,][]{2021ATel14609....1X, 2022MNRAS.511.3922J, 2022ApJ...926L..10M}, as is standard for dipping sources
\citep[see e.g.,][]{1997xrb..book.....L, 1998ApJ...494..753K}. This would imply a close binary system. As such, the companion star will be heavily irradiated by X-rays from the inner accretion disk of the black hole. Such heavy irradiation could drive a strong stellar wind even from a low mass companion star similar to the process of ``ablation'' in neutron star low-mass X-ray binaries (LMXBs) --- a process by which a companion star's outer layer is liberated and depleted through X-ray irradiation and a pulsar wind from the compact object \citep[see e.g.,][]{2023MNRAS.520.3416K}. X-ray absorption by clumps of such liberated partially-ionized material could be responsible for the dips. 

\citet{1991Natur.350..136P} showed that an irradiating X-ray flux of $\sim4\times10^{11}\,\mathrm{erg\,cm^{-2}\,s^{-1}}$ from a neutron star in an interacting LMXB system will cause the companion star in such a system to expand by a factor of 2 to 4 and consequently lose mass. Assuming the orbital period $P$ of the \rm{MAXI~J1803-298} system to be $7\,\mathrm{hours}$, if the black hole and the companion star have masses $M_{\rm{BH}}$ and $M_{\rm{CS}}$ respectively, then following Kepler's law, the separation $a$ between them is
\begin{equation}
a=\left[\frac{G(M_{\rm{BH}}+M_{\rm{CS}})}{(2\pi)^{2}}P^{2}\right]^{1/3},    
\label{eqn:two}
\end{equation}
where $G$ is the gravitational constant. The X-ray flux irradiating the companion star $F_{X}^{ir}$ is related to the observed X-ray flux $F_{X}^{ob}$ by the equation;
\begin{equation}
    F_{X}^{ir}\approx F_{X}^{ob}\left(\frac{d}{a}\right)^{2},
\label{eqn:three}    
\end{equation}
where $d$ is the distance to the system. Using the best-fit model to the joint spectra of Epoch 1, the unabsorbed $0.1-300\,\mathrm{keV}$ flux is $\sim10^{-8}\,\mathrm{erg\,cm^{-2}\,s^{-1}}$. From Equations \ref{eqn:two} and \ref{eqn:three}, if we assume the total mass of the system (i.e. $M_{\rm{BH}}$+$M_{\rm{CS}}$) to be $10M_{\odot}$ and at a distance of $8\,\mathrm{kpc}$ away \citep[e.g.,][]{2022ApJ...926L..10M}, the irradiating flux will be of the order $F_{X}^{ir}\approx10^{14}\,\mathrm{erg\,cm^{-2}\,s^{-1}}$. This value is significantly higher than the predicted threshold from \citet{1991Natur.350..136P}, further supporting the case for ablation as a likely origin for the recurring dips, with the caveat that \rm{MAXI~J1803-298} is a black hole and not a neutron star. 

While it is harder to infer the presence of ablated material in the absence of eclipses from the companion star, detailed phase-resolved spectroscopy from observations covering a significant number of dipping cycles in BHXBs will be critical to confirming if indeed ablation also occurs in BHXBs as in neutron star LMXBs.
 
An absorption line is significantly detected ($>4\sigma$) at $6.61\pm{0.05}\,\mathrm{keV}$ in the dip spectra.
This line is coming from the material of the obscurer that is plausibly moving at a relatively slow speed and in a direction away from our line of sight. This further confirms that the material of the obscurer responsible for the dips is not the same as that in typical outflowing winds. 

\subsection{Epoch 2: Intermediate State} \label{subsec:four-two}
Epoch 2 caught \rm{MAXI~J1803-298} in the intermediate state near the peak of the outburst. The source displayed extreme short-term variability during this epoch,
strong relativistic disk reflection signatures and an absorption line at $6.86^{+0.11}_{-0.08}\,\mathrm{keV}$ (detected at $\sim3\sigma$). Associating the absorption line with {Fe~\sc{xxv}} would imply a fast wind outflow velocity of $\sim7200\,\mathrm{km\,s^{-1}}$. Fitting the reflection spectra with the \texttt{relxillCp} flavor of the \texttt{relxill} family of models adequately reproduced the spectra. The model also confirms the source to have a high inclination of $i=68^{+5}_{-4}\,\mathrm{\degree}$ and to be rapidly spinning with $a_*=0.98^{+0.01}_{-0.02}$. These values are consistent within errors to those reported in \citet{2023ApJ...949...70C}. Those authors also showed that the power-law normalization drives the variability seen in the source during this epoch, implying that changes in the corona rather than the disk may play a more significant role in the observed variability. 

\subsection{Epoch 3: Intermediate State} \label{subsec:four-three}
The broadband spectrum during Epoch 3 shows strong relativistic reflection (see Fig. \ref{fig:mo_refl}).
Because the source is in the intermediate state during Epoch 3 and reaches the soft state soon after, the flux increase observed here (see Fig. \ref{fig:lc_nustar-nicer}) may point to the gradual filling of the inner disk as the source transitions into the soft state. The consistently lower value of the \texttt{diskbb} normalization for the high flux compared to the low flux spectra is also expected in such a scenario since the normalization is defined as;
\begin{equation}
    Norm=\left(\frac{r_{in}}{D_{10}}\right)^{2}cos\theta,
\label{eqn:four}
\end{equation}
where $r_{in}$ is the ``apparent'' disk inner radius in $\mathrm{km}$, $D_{10}$ is the distance to the source in units of $10\,\mathrm{kpc}$ and $\theta$ is the inclination of the inner disk. $r_{in}$ is related to the true inner radius $R_{in}$ by the equation $R_{in}=\Xi f^{2}r_{in}$ ($f$ and $\Xi$ are the spectral hardening factor and the relativistic correction factor, respectively \citealp{1995ApJ...445..780S}). It is thus plausible that the increased flux seen in the light curve during Epoch 3 (i.e. the high flux interval) is  caused by an increase in the soft disk photon flux. This is supported by the model-independent HID of Fig. \ref{fig:hid_nu} which shows that soft photons dominate the higher count-rate regime while hard photons tend to dominate the lower count-rate regime. The disk temperature is higher during the high flux state as expected if the high flux spectra are linked to the inner regions of a filling accretion disk. 
This supports the canonical picture of BHXB state evolution, where the accretion disk is truncated in the hard state at the onset of an outburst, gradually moving inward as the outburst evolves through the intermediate states and reaches the ISCO in the soft state \citep[e.g.,][]{2007A&ARv..15....1D}.

If we take $f$ and $\Xi$ to be same for both the high and the low flux spectra and assume the difference in \texttt{diskbb} normalization to be purely driven by the disk inner edge extending inwards towards the ISCO, we can estimate how much the inner radius has changed over this interval. From the relation in Equation \ref{eqn:four} we can have
\begin{equation}
    \frac{R_{in}^{HF}}{R_{in}^{LF}} = \left(\frac{Norm_{HF}}{Norm_{LF}}\right)^{1/2},
\label{eqn:five}
\end{equation}
where $R_{in}^{HF}$ and $R_{in}^{LF}$ are the inner radii of the high and low flux spectra, respectively. This gives $R_{in}^{HF}\sim 0.88R_{in}^{LF}$ implying that during the duration of increased flux, the disk inner radius $R_{in}$ has dropped to about 88 percent of its earlier value.

A more extreme version of this behavior, happening on much shorter timescale, is reported for \rm{GRS~1915+105} based on \textit{RXTE} observations \citep{1997ApJ...488L.109B} where rapid light curve variability and the associated spectral changes were attributed to the rapid disappearance of the inner region of an accretion disk, followed by a slower refilling of the emptied region, caused by viscous-thermal instability. \citet{1997ApJ...488L.109B} showed that during bursts (or high flux epochs), the temperature rises while the inner radius decreases and during quiescent (or low flux) phases, the temperature drops while the inner radius increases. They linked this to a factor of 2 increase in accretion rate during the bursts compared to the quiescent phases. It is worth mentioning that this interpretation of the observed spectral variability in {\rm GRS~1915+105} is not unique, and is caveated by the limited low energy pass band as well as energy resolution of \textit{RXTE}. For example, using an \textit{RXTE} observation of {\rm GRS~1915+104} while in the low-hard state, \citet{2001A&A...370L..17V} showed that episodes of significant flux decrease seen in the data correspond to a drop mostly in the $8-25\,\mathrm{keV}$ count rate. They interpret this as the ejection of a Compton cloud.

We posit that the large amplitude change in flux observed during Epoch 3 indicates replenishing of the inner accretion disk initiated by a viscous-thermal instability. During the low flux phase, the inner part of the disk is empty (or truncated) and slowly fills via steady accretion. The surface gravity increases as each annulus moves towards the unstable point in the $\dot{M}-\Sigma$ plane on the viscous timescale ($\dot{M}$ is the local accretion rate and $\Sigma$ is the surface density). However, as one of the annuli reaches an unstable point, it experiences a significant increase in $\dot{M}$ resulting in a chain reaction that switches on the inner disk. This causes increased flux and a hotter radius with smaller $R_{in}$ \citep[see e.g.,][]{1974ApJ...187L...1L, 1976MNRAS.175..613S, 1988ApJ...332..646A, 1991A&A...249..574L, 1995ApJ...443L..61C}.

The discussion presented here is based on the assumption of a constant spectral hardening factor $f$. The effect of a change in $f$ through the disk atmosphere between the low and the high flux spectra has not been considered which, if significant, might be important in determining the exact change in $R_{in}$ between the low and the high flux intervals \citep[see e.g.,][]{2013MNRAS.431.3510S, 2020ApJ...890...53S}.

The temperature of the corona is fairly well constrained during this epoch to be $kT_{e}=53^{+44}_{-9}\,\mathrm{keV}$. This is consistent with the behavior of BHXBs prior to transitioning to the canonical soft state 
as studies of the evolution of the coronal cutoff energy show that the spectral turnover at high energies disappears after transition to the soft state \citep[e.g.,][]{2008ApJ...679..655J, 2009MNRAS.400.1603M}. We therefore posit that this observation caught the source in the process of transitioning into the soft state, providing a glimpse into state transition.

\subsection{Epoch 4: Soft State} \label{subsec:four-four}
\rm{MAXI~J1803-298} has transitioned into the soft, disk-dominated state during Epoch 4, with the power law contributing only $\sim5\%$ to the total $0.1-100\,\mathrm{keV}$ flux. 
While the \texttt{relxillCp} flavor of the \texttt{relxill} family of models reproduces the overall reflection spectra adequately, it struggles to constrain important spectral parameters such as the spin, inclination and the inner emissivity index.

Replacing \texttt{relxillCp} with \texttt{relxillNS} yields better overall fit statistics and improves constraints on the spectral parameters. 
Because the \texttt{relxillNS} model was developed to describe reflection from disks around neutron stars, it adopts a single-temperature thermal blackbody as its irradiating continuum, rather than a multi-temperature disk blackbody. The single-temperature approximation is sufficient for our purposes since the light-bending effect that causes disk photons to self-irradiate will be most important in the innermost regions of the accretion disk. 

Using \textit{RXTE} observations of \rm{XTE~J1550-564} in the very soft state, \citet{2020ApJ...892...47C} compared fits from \texttt{relxillCp} with \texttt{relxillNS} and a few other models and found that \texttt{relxillCp} tends to overfit the spectrum at high energies --- because of the softness of the spectrum --- and struggles to simultaneously capture broad iron line features while maintaining an appropriate fit to the overall spectrum. The authors concluded that only \texttt{relxillNS} is capable of providing a good overall fit and, at the same time, capture the subtleties of the Fe K region. Figure \ref{fig:epoch_4_mod} (left) reveals a related tendency for the \texttt{relxillCp} fit to the soft state data of \rm{MAXI~J1803-298}, it tends to fit for the spectrum above $\sim10\,\mathrm{keV}$ over the continuum.

\citet{2022MNRAS.514.3965D} showed, consistent with previous estimates, that for a maximally spinning black hole with a compact primary source of radiation close to the black hole, returning radiation can make up $40-80\%$ of the total observed flux and for a spin of $0.95$, returning radiation can contribute at most $30\%$ to the total flux.
The reflection component contributes $\sim21\%$ of the total flux in the \texttt{relxillNS} fit to Epoch 4 spectra.

While the application of \texttt{relxillNS} here is only a first approximation, disk self-irradiation due to returning radiation is a likely cause of the reflected component seen in the very soft state of BHXBs when the system is disk-dominated.

This epoch also suggests the presence of plausible disk winds from the prominent spectral absorption line detected at $6.74^{+0.05}_{-0.06}\,\mathrm{keV}$. This corresponds to an outflow velocity of $1800^{+2200}_{-2700}\,\mathrm{km\,s^{-1}}$ if associated with {Fe~\sc{xxv}}. This is consistent with typical disk wind velocities in BHXBs which are known to be less than $1000\,\mathrm{km\,s^{-1}}$ \citep[e.g.,][]{2006Natur.441..953M, 2006ApJ...646..394M}.
However, there have been instances where extremely fast outflows, with velocities greater than $\sim9000\,\mathrm{km\,s^{-1}}$, were inferred \citep[see e.g.,][]{2012ApJ...746L..20K, 2012MNRAS.425.2436C, 2014ApJ...784L...2K, 2021ApJ...906...11W, 2021MNRAS.508..475C}. 

\section{Conclusion} \label{sec:five}
We probed the X-ray spectral evolution of the BHXB \rm{MAXI~J1803-298} using \textit{NuSTAR} and \textit{NICER} data from its 2021 outburst. Our main conclusions are summarized as follows:

\begin{itemize}
    \item Relativistic reflection modeling indicates the source is rapidly spinning ($a_*=0.990\pm{0.001}$) and observed close to the disk plane ($i=70\pm{1}{\degree}$).
    \item The source showed flux dips in its light curves while in the hard/intermediate states. We attribute these to photo-electric absorption from moderately ionized obscuring material coming into the line-of-sight, most likely linked to the companion star. The material of the obscurer might be moving in a direction opposite to our line-of-sight, based on the velocity shift of the Fe K absorption line detected in its spectra.
    \item Spectral absorption lines plausibly from iron were also present in the intermediate and the soft states. These lines are indicative of moderate to extreme winds from the outer accretion disk of the black hole.
    \item A flux rise seen during one of the epochs in the intermediate state is found to be dominated by soft disk photons and is believed to signal the filling of the inner accretion disk towards ISCO as the source transitions into the soft state.
    \item While in the soft state, disk self-irradiation is plausibly responsible for most of the reflection features observed in the source.
\end{itemize}
Higher spectral resolution data \citep[see e.g.,][]{2022NatAs...6.1364G} would be crucial to putting a better constraint on the velocity shifts and the ionization states of any outflowing winds.

\begin{acknowledgments}
The authors would like to thank the anonymous referee as well as Poshak Gandhi and Brian Grefenstette for comments that improved the clarity of the manuscript. This work was supported under NASA Contract No. NNG08FD60C, and made use of data from the \textit{NuSTAR} mission, a project led by the California Institute of Technology, managed by the Jet Propulsion Laboratory, and funded by the National Aeronautics and Space Administration. AI acknowledges support from the Royal Society. JAT acknowledges partial support from the \textit{NuSTAR} Guest Observer program under NASA grant 80NSSC22K0059. This research has made use of the \textit{NuSTAR} Data Analysis Software (NuSTARDAS) jointly developed by the ASI Space Science Data Center (SSDC, Italy) and the California Institute of Technology (Caltech, USA).
\end{acknowledgments}

%

\vspace{5mm}
\facilities{\textit{NuSTAR}, \textit{NICER}, \textit{MAXI}}

\software{
          XSPEC \citep{1996ASPC..101...17A}, 
          XSTAR \citep{2001ApJS..133..221K},
          relxill \citep{2014MNRAS.444L.100D, 2014ApJ...782...76G}          
          }

\bibliography{bibtex}{}
\bibliographystyle{aasjournal}



\end{document}